\documentclass{aa}

\usepackage{amsmath}
\usepackage{subfigure}
\usepackage{graphicx}
\usepackage{nicefrac}
\usepackage{xcolor}
\usepackage{lipsum}
\usepackage{caption}
\usepackage{dblfloatfix}    
\usepackage{txfonts}
\usepackage[hidelinks]{hyperref}

\usepackage{soul}

\newcommand{\hmpc}{h^{-1}\mathrm{Mpc}}

\begin{document}

   \title{Galaxy clustering multi-scale emulation}

   \author{Tyann Dumerchat, Julian Bautista
          \inst{\ref{amu}}}

   \institute{Aix Marseille Univ, CNRS/IN2P3, CPPM, Marseille, France \\
              \email{dumerchat@cppm.in2p3.fr}
              \label{amu}
             }


 
  \abstract{Simulation based inference has seen increasing interest in the past few years as a promising approach to model the non linear scales of galaxy clustering. 
  The common approach using Gaussian process is to train an emulator over the cosmological and galaxy-halo connection parameters independently for every scales. We present a new Gaussian process model allowing to extend the input parameter space dimensions and to use non-diagonal noise covariance matrix. 
  We use our new framework to emulate simultaneously every scales of the non-linear clustering of galaxies in redshift space from the \textsc{AbacusSummit} N-body simulations at redshift $z=0.2$. The model includes nine cosmological parameters, five halo occupation distribution (HOD) parameters and one scale dimension. Accounting for the limited resolution of the simulations, we train our emulator on scales from $0.3~\hmpc$ to $60~\hmpc$ and compare its performance with the standard approach of building one independent emulator for each scales. The new model yields more accurate and precise constraints on cosmological parameters compared to the standard approach.
  As the new model is able to interpolate over the scales space, we are also able to account for the Alcock-Paczynski distortion effect leading more accurate constraints on the cosmological parameters.}

   \keywords{Cosmology -- Large scale structures --
                Machine learning -- Computational methods
               }

   \maketitle
%

\section{Introduction}

Galaxies are one of the best tracers of the underlying dark matter distribution in our Universe as a function of time. 
Spectroscopic galaxy surveys have become a powerful probe of the physical laws governing the formation of the large-scale structures, since they provide precise spatial information for millions of galaxies over large volumes.
Surveys such as the 
2dFGRS \citep{colless2dFGalaxyRedshift2003}, 
6dFGS \citep{jones6dFGalaxySurvey2004}, 
WiggleZ \citep{drinkwaterWiggleZDarkEnergy2010}, 
VIPERS \citep{guzzoVIMOSPublicExtragalactic2014}, 
SDSS BOSS and eBOSS \citep{eisensteinSDSSIIIMassiveSpectroscopic2011, 
blantonSloanDigitalSky2017, dawsonBaryonOscillationSpectroscopic2013,  
dawsonSDSSIVExtendedBaryon2016}, have been (and will continue being) one of the strongest tools to understand the accelerated expansion of our Universe and test models of dark energy or modified theories of gravity \citep{weinbergObservationalProbesCosmic2013, zhaiEvaluationCosmologicalModels2017, hutererGrowthCosmicStructure2023a, houCosmologicalProbesStructure2023}.

The analysis of spectroscopic galaxy surveys consists in the measurement and modelling of two main physical features present in the two-point function of the density field: the baryon acoustic oscillations (BAO) and the redshift space distortions (RSD). 
While the BAO can be well modelled by linear theory, since they manifest on correlations at large linear scales (around $100 \hmpc$), RSD are more challenging since there is valuable information on small  mildy non-linear scales (below few tens of $\hmpc$), therefore harder to model accurately. The latest measurements of BAO and RSD from SDSS have put the tighest constraints on the growth history to date \citep{alamCompletedSDSSIVExtended2021, rossClusteringSDSSDR72015, howlettClusteringSDSSMain2015, alamClusteringGalaxiesCompleted2017, bautistaCompletedSDSSIVExtended2021,gil-marinCompletedSDSSIVExtended2020,demattiaCompletedSDSSIVExtended2021,tamoneCompletedSDSSIVExtended2020, houCompletedSDSSIVExtended2021, neveuxCompletedSDSSIVExtended2020, dumasdesbourbouxhelionCompletedSDSSIVExtended2020}. 
Ongoing and future galaxy surveys such as Euclid \citep{laureijsEuclidDefinitionStudy2011,euclidcollaborationEuclidPreparationEuclid2022} and the Dark Energy Spectroscopic Instrument (DESI, \citealt{desi_collaboration_desi_2016, desicollaborationOverviewInstrumentationDark2022})
 span deeper and wider fields of view and will measure tens of millions of galaxies. 
 Such samples of galaxies will significantly reduce statistical uncertainties in the two-point functions, requiring higher precision clustering models to derive unbiased cosmological constraints.
 
Standard perturbative theories attempt to model mildly non-linear scales analytically 
\citep{taruya_baryon_2010,reid_towards_2011,carlson_convolution_2013,wang_analytic_2014}. 
Even more modern approaches (including loop corrections) such as the effective field theory of large scale structure (EFToLSS, \citealt{pajer_renormalization_2013,carrasco_effective_2014}) try to push the model validity to smaller scales, but ultimately such perturbative approaches are limited as the density perturbations on small scales become too large, especially at low redshifts.


An alternative to perturbation theory is to build a simulation-based model 
to extract information from the small scale clustering. 
A recent approach consists in running several N-body simulations for 
different cosmological models, compute the summary statistics for each and 
use them to predict the clustering for a new set of cosmological parameters. 
This process is commonly referred to as emulation.
Brute iterative solving of dynamical equations in N-body simulations provide the high fidelity description of the matter non-linear clustering, although they are computationally expensive to produce. 
In the past years several suites of N-body simulations have been produced for this purpose \citep{derose_aemulus_2019,villaescusa-navarro_quijote_2020, maksimova_abacussummit_2021}.
Several emulators were developed \citep{kobayashiAccurateEmulatorRedshiftspace2020} and some of these were applied to real data to extract cosmological parameters from real data \citep{kobayashiFullshapeCosmologyAnalysis2022, chapmanCompletedSDSSIVExtended2022}.

\begin{figure*}
	\centering
	\includegraphics[width=0.9\textwidth]{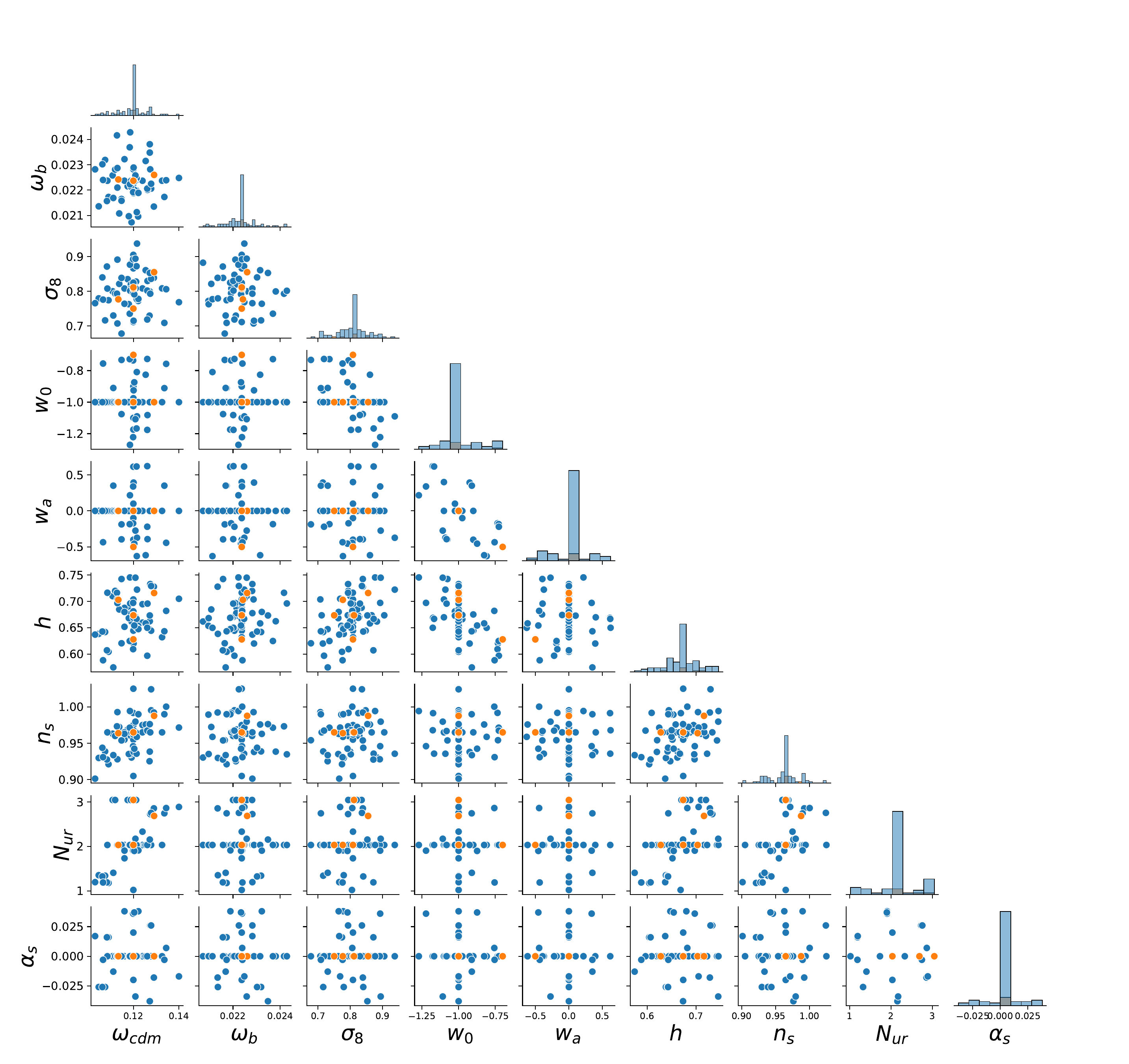}
	\caption{Sampling over the 9 dimensional cosmological space around Planck2018 $\Lambda$CDM. In blue the of 88 cosmologies that used to train the emulator, in orange the 6 cosmologies used as a test set. The dark matter density $\rm \omega_{cdm}$, the baryon density $\rm \omega_{b}$, the clustering amplitude $\sigma_8$, the dark energy equation of state $w(a) = w_0 + w_a(1-a)$, the Hubble constant h, the spectral tilt $\rm n_s$, the density of massless relics $\rm N_{ur}$ and the running of the spectral tilt $\alpha_s$ \citep{maksimova_abacussummit_2021}.}
	\label{fig:cosmogrid}
\end{figure*}

Any statistics derived from the properties of dark matter tracers suffers from cosmic variance.
This effect decreases as the volume of the simulation increases, but ultimately every measured observable comes with its error budget.
Hence, an emulator should consider the uncertainty of the training set along with the uncertainty of the predictions. A popular emulation algorithm used in previous works for galaxy clustering \citep{yuan_stringent_2022, zhai_aemulus_2022, kwan_galaxy_2023} is the Gaussian Process (GP).
This Bayesian machine learning method with physical interpretation not only provides a prediction for the model but also an associated uncertainty.
However it is well known that GP models scale poorly with the number of training points $N$ as it requires to build and inverse $N \times N$ matrix (N is the number of training samples) a lot of times during the training phase. Each of these operations costs $O(N^2)$ memory and $O(N^3)$ time.
As a consequence the dimension of the input space is often restricted to the cosmology and galaxy-halo connection parameters.
Particularly the standard approach of the previous works is to construct an independent emulator for each separation or Fourier mode, ignoring the well-known correlations between scales.
In this work we propose a new Gaussian process model, allowing to extend the input parameter space, thus to interpolate/extrapolate over a wider range of parameters such as scales, redshift bins, selection effects etc.
As a prove of concept, we built an emulator for the two point correlation function from a redshift $z=0.2$ snapshot of the \textsc{AbacusSummit} suite \citep{maksimova_abacussummit_2021}, and extended the input parameter space to include the different separation scales. We implemented our new framework on GPU with the \textsc{Jax} library.

This paper is organised as follows. In section \ref{sec:data} we present the \textsc{AbacusSummit} N-body simulations, the galaxy-halo connection and the observable. In section \ref{sec:ML} we describe the Gaussian process framework along with our new model. In section \ref{sec:emu_perf} we validate our emulator performance compare to a standard model. Finally, in section \ref{sec:reco} we explore the constraining power of our model.



\section{Emulating clustering from N-body simulations}
\label{sec:data} 

In this section we describe the suite of N-body simulations used in this work. Since these are dark matter-only, we use halo occupation distribution (HOD) to populate galaxies. Then, we discuss the summary statistics we use to evaluate the performance of our emulation technique.

\subsection{The \textsc{AbacusSummit} suite of simulations}
\label{sec:data:abacus}

\textsc{AbacusSummit} \citep{maksimova_abacussummit_2021} is a suite of N-body simulation ran with the Abacus N-body code \citep{garrison_abacus_2021} on the Summit supercomputer of the Oak Ridge Leadership
Computing Facility. These simulations were designed to match the requirement of the Dark Energy Spectroscopic Survey (DESI). It is composed of hundreds of periodic cubic boxes evolved in comoving coordinates to different epochs from redshift $z=8.0$ to $z=0.1$. 
The base simulations consist of $6912^3$ particles of mass $2\times 10^9 ~ h^{-1} M_\odot$ in a $2~ h^{-1}\mathrm{Gpc})^3$ volume, yielding to a resolution of about $0.3~\hmpc$. The suite provides halo catalogs formed by identifying groups of particles, built using the specialized spherical overdensity
based halo finder \textsc{CompaSO} (see \citealt{hadzhiyska_textsccompaso_2021} for more details).

\textsc{AbacusSummit} spans several cosmological models sampled around the best-fit $\Lambda$CDM model to Planck 2018 data \citep{planck_collaboration_planck_2020} with 9 different parameters $\theta_{\Omega} = \{\omega_\mathrm{cdm}, \omega_\mathrm{b},\sigma_8,w_0,w_a,h,n_\mathrm{s}, N_\mathrm{ur}, \alpha_\mathrm{s} \}$: 
the dark matter density $\omega_\mathrm{cdm}$, the baryon density $\omega_\mathrm{b}$, the clustering amplitude $\sigma_8$, the dark energy equation of state $w(a) = w_0 + w_a(1-a)$, the Hubble constant $h$, the spectral tilt $n_\mathrm{s}$, the density of massless relics $N_\mathrm{ur}$ and the running of the spectral tilt $\alpha_\mathrm{s}$. A flat curvature is assumed for every cosmology. 

Figure \ref{fig:cosmogrid} shows the sampling over the 9 dimensional space of 88 cosmologies (in blue) that will be used to train the emulator, and 6 (in orange) that will be used as a test set all with the base resolution. 
The test set includes the center of the hypercube: Planck2018 $\Lambda$CDM cosmology with 60~meV neutrinos, the same cosmology with massless neutrinos, 4 secondary cosmologies with low $\omega_\mathrm{cdm}$ choice based on WMAP7, a $w$CDM choice, a high $N_\mathrm{ur}$ choice, and a low $\sigma_8$ choice. The train set provides a wider coverage of the parameter space. In the following we call $X_{\Omega} $ the $88 \times 9$ matrix containing the cosmological parameters of the training set.

All simulations were run with the same initial conditions so changes in the measured clustering are only due to  changes in cosmology.
To test the cosmic variance, \textsc{AbacusSummit} also provides an other set of 25 realisations with same cosmology and different initial conditions and an additional set of 1400 smaller $\rm(500~\hmpc)^3$ boxes with same cosmology, mass resolution but different phases.
The small boxes are used to compute sample covariance matrices of our data vectors. All those simulations were run with the base Planck2018 cosmology. 
While snapshots are available for several redshifts,  we only use snapshots at redshift $z=0.2$ throughout this work\footnote{While a $z=0.1$ snapshot exists for AbacusSummit, the smaller boxes for covariance matrix estimations were not available.}.

\subsection{Halo occupation distribution model}
\label{sec:data:hod}

We use the halo occupation distribution (HOD, \citealt{zheng_galaxy_2007}) framework to assign galaxies to dark-matter halos of the simulations. 
This framework assumes that the number of galaxies  $N$ in a given dark matter halo follows a probabilistic distribution $P(N|\Omega_h)$, where $\Omega_h$ can be a set of halo properties. In this work we assume a standard vanilla HOD, where this distribution only depends on the halo's mass $\Omega_h = M$. Moreover, the occupation is decomposed into the contribution of central and satellite galaxies $\langle N(M) \rangle = \langle N_{\rm cen}(M) \rangle + \langle N_{\rm sat}(M) \rangle$. 
The number of central galaxy occupation follows a Bernoulli distribution while the satellite galaxy occupation follows Poisson distribution. The mean of these distributions are defined as : 

\begin{equation}
\langle N_{\rm cen} \rangle = \frac{1}{2} {\rm erfc}\left( \frac{{\rm log_{10}}(M_{\rm cut}/M)}{\sqrt{2}\sigma}\right),
\label{eq:hod_cen}
\end{equation}
\begin{equation}
\langle N_{\rm sat} \rangle = \left [ \frac{M - \kappa M_{\rm cut}}{M_1} \right]^{\alpha}\langle N_{\rm cen} \rangle.
\label{eq:hod_sat}
\end{equation}

With erfc($x$) the complimentary error function and $\theta_{\rm hod} = \{M_{\rm cut}, \sigma, \kappa, M_1,\alpha \}$ the HOD parameters : $M_{\rm cut}$ the minimum mass to host a central galaxy, $\sigma$ the width of the cutoff profile (i.e the steepness of the transition 0 to 1 in the number of central galaxy), $\kappa M_{\rm cut}$ the minimum mass to host a satellite galaxy, $M_1$ the typical mass to host one satellite galaxy, and $\alpha$ the power law index of the satellite occupation. 

\begin{figure}
	\centering
	\includegraphics[width=1.\columnwidth]{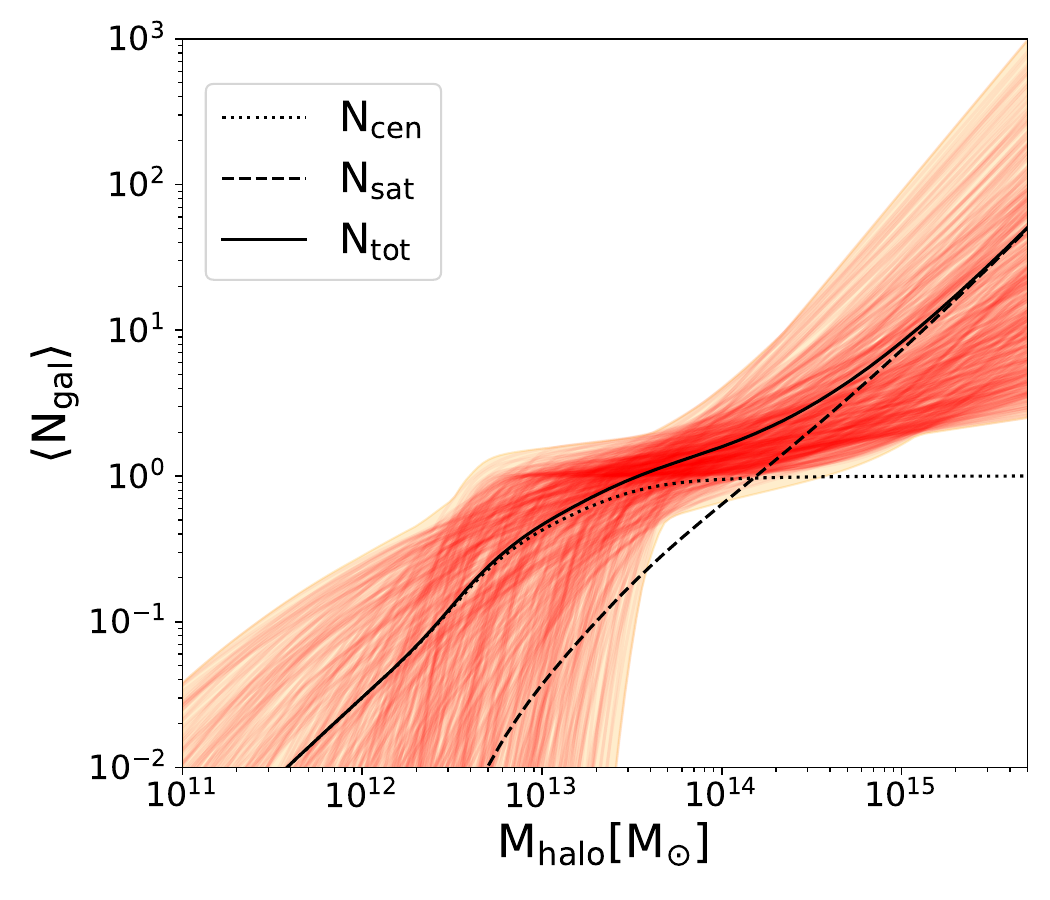}
	\caption{Different forms that takes the total occupation of galaxy for the 600 HOD parametrisation used. The black lines show the mean HOD configuration for $\langle N_{\rm cen}(M) \rangle$, $\langle N_{\rm sat}(M) \rangle$ and $\langle N_{\rm tot}(M) \rangle$. }
	\label{fig:hod_mean}
\end{figure}

\begin{table}
\caption{Cosmological and HOD parameter names and ranges sampled to build the emulator training and testing sets. The cosmology bounds come from the
smallest hypercube around the AbacusSummit cosmology grid. The HOD bounds come from the the minimum and maximum in each parameter within the 600 HOD.}
\centering
\begin{tabular}{ccc}
\hline
\hline
  & 
  Parameter &
  Range \\
  \hline

Cosmology & $\rm \omega_{cdm}$& [0.103 ,0.140] \\
 & $\rm \omega_{b}$& [0.0207,0.0243 ] \\
& $\sigma_8$& [ 0.678,0.938] \\
& $\rm w_0$& [-1.271,-0.726]\\
& $\rm w_a$& [-0.628,0.621] \\
& $\rm h$& [  0.575 , 0.746 ] \\
& $\rm n_s$& [ 0.901, 1.025] \\
& $\rm N_{ur}$& [1.020 , 3.046 ]\\
& $\rm \alpha_s$& [-0.038 , 0.038] \\

\hline

HOD &$\rm \alpha$& [0.30, 1.48] \\
& $\rm \kappa$&  [0.00, 0.99]\\
 & $\rm log_{10}M_1$&  [13.6, 15.1]\\
 & $\rm log_{10}M_{cut}$& [12.5, 13.7]  \\
 & $\rm log_{10}\sigma$& [-1, 0] \\

\hline
\hline
\end{tabular}
\label{tab:cosmo_hod_range}
\end{table}

We use the \textsc{AbacusHOD} implementation  \citep{yuan_abacushod_2022} where for each halo, if a random number - drawn from a normal distribution between 0 and 1 - is smaller than $\langle N_{\rm cen}\rangle$, a central galaxy is assigned to the center of mass of the halo with the same velocity. Satellite galaxies do not assume a fixed distribution profile such as the Navarro-Frenk-White profile (\cite{navarro_universal_1997}), but rather follow the distribution of the dark matter particles. Specifically, each particle inside an halo has the same probability of hosting a satellite galaxy $p =\langle  N_{\rm sat}\rangle/N_p$ with $N_p$ the number of particles in a given halo. For every particle, a random number between 0 and 1 is drawn, if it is smaller than $p$ a galaxy is assigned to their position with the same velocity. The \textsc{abacusutils} package implements this halo occupation scheme and it is publicly available\footnote{\url{https://github.com/abacusorg/abacusutils}}. 


To build our trainset, we populate each of the 88 boxes with 600 HODs selected using a latin hypercube sampling, resulting in 52800 different clustering measurements. For the test set, we use a different set of 20 HODS to populate the 6 test cosmologies.
We chose the parameters range according to the bright galaxy sample of the DESI one percent survey \citep{prada_desi_2023}.
In the following we call $X_{\rm HOD} $ the $600 \times 5$ matrix containing the HOD parameters of the training set.
Note that using the same HOD sampling for every cosmology results in a grid-structured parameter space $X_{\Omega} \otimes X_{\rm HOD} $ where $\otimes $ is the Kronecker product. This peculiar data structure will be usefull to build the emulator in section \ref{sec:ML}. 
Figure \ref{fig:hod_mean} illustrates the different forms that the total occupation of galaxy takes for the wide range of HOD parametrisation that we use. The black lines show the mean HOD configuration for $\langle N_{\rm cen}(M) \rangle$, $\langle N_{\rm sat}(M) \rangle$ and $\langle N_{\rm tot}(M) \rangle$. In table \ref{tab:cosmo_hod_range} we summarize the sampling range of every cosmological and HOD parameters.

\subsection{Observable of interest: the correlation function}
\label{sec:data:observable} 

To capture the cosmological information in the small scales clustering of galaxies we use as an observable the standard statistic called two point correlation function (2PCF). 
It is defined as $\xi(\textbf{r}_1-\textbf{r}_2) = \langle \delta_g(\textbf{r}_1)\delta_g(\textbf{r}_2) \rangle$ with $\delta_g(\textbf{r})$ the over density of galaxy at position $\textbf{r}$. The 2PCF is measured in a catalog by counting the pairs of galaxies with the well known Landy-Szalay estimator \citep{landy-szalay}:
\begin{equation}
\xi = \frac{DD -DR - RD + RR}{RR},
\label{eq:LS}
\end{equation}
where $DD$, $RR$, $DR$ and $RD$ are the normalized pair counts of data-data, random-random, data-random and random-data catalogs. The random catalog is used to define the window function of the survey.
Because each galaxy has its own peculiar velocity on top of the Hubble flow, the measured positions of galaxies along the line of sight (LoS) are modified by the so called  redshift-space distortions (RSD) as :
\begin{equation}
d_s = d + \frac{v_{\parallel}}{H(z)}(1+z),
\label{eq:rsd}
\end{equation}
with $d$ and $d_s$ the true and distorted radial position, $v_{\parallel}$ the peculiar velocity along the LoS, and $H(z)$ the Hubble parameter at that redshift. Note that because we are using snapshots of N-body instead of light cones, the redshift evolution is not simulated and here $z$ has the same value of 0.2 for all galaxies.

Because of RSD, the 2PCF loses its isotropy and no longer only depends on the relative separation between the galaxy pairs $s = |\textbf{s}_1 - \textbf{s}_2|$, but on the triangles formed by each pair of galaxies and the observer. Hence the pair counts have to be binned in three dimensions ($s,\mu,\theta$), with $\mu$ the cosine of the angle between the observer LoS (passing through the midpoint between the galaxies) and the separation vector of the galaxy pair, and $\theta$ the angle between the two individual lines of sight. 
The wide-angle dependence is minor for distant surveys, so we can perform, with a minor loss of information, the projection according to \cite{yoo_wide_2015}:
\begin{equation}
\xi(s,\mu) = \frac{\int \xi(s,\mu,\theta)P(s,\mu,\theta)d\theta}{\int P(s,\mu,\theta) d\theta},
\label{eq:wideangle_proj}
\end{equation}
where $P(s,\mu,\theta)$ is a probability distribution proportional to the number of pairs at a given triangle configuration and is survey dependant. Note that the wider the distribution in $\theta$ is, the more this projection makes the observable survey dependant. 
This is equivalent to straightly binning the pair counts in $(s,\mu)$.

The 2PCF can then be decomposed as a series of Legendre polynomials :

\begin{equation}
\xi(s,\mu) = \sum_l P_l(\mu) \xi_l(s),
\label{eq:legendre1}
\end{equation}
with
\begin{equation}
\xi_l(s) = \frac{2l+1}{2}\int \xi(s,\mu) P_l(\mu) d\mu .
\label{eq:legendre2}
\end{equation}

As this work is a proof of concept for our new emulation strategy, we will focus on the monopole $\xi_0$ hereafter. The study of all multipoles is left for future work. The most expensive part of computing the estimator defined in Eq. \ref{eq:LS} is the random-random pair counting. As we have cubic boxes with periodic boundary conditions, $RR$ can be analytically computed with a fixed LoS for every pair counts and we can make use of the natural estimator  \cite{peebles} :

\begin{equation}
\xi = \frac{DD}{RR} - 1,
\label{eq:peebles}
\end{equation}
greatly reducing computational time. 
 This methodology could however bias the clustering as a fixed LoS is equivalent to assuming $\theta = 0$ for each pair of galaxies, or that the observer is infinitely far away. 
This flat sky approximation should break down for low-z surveys and large separation bins. We tested this hypothesis with the 25 Planck2018 boxes, using the median HOD.

\begin{figure}
	\centering
	\includegraphics[width=1.\columnwidth]{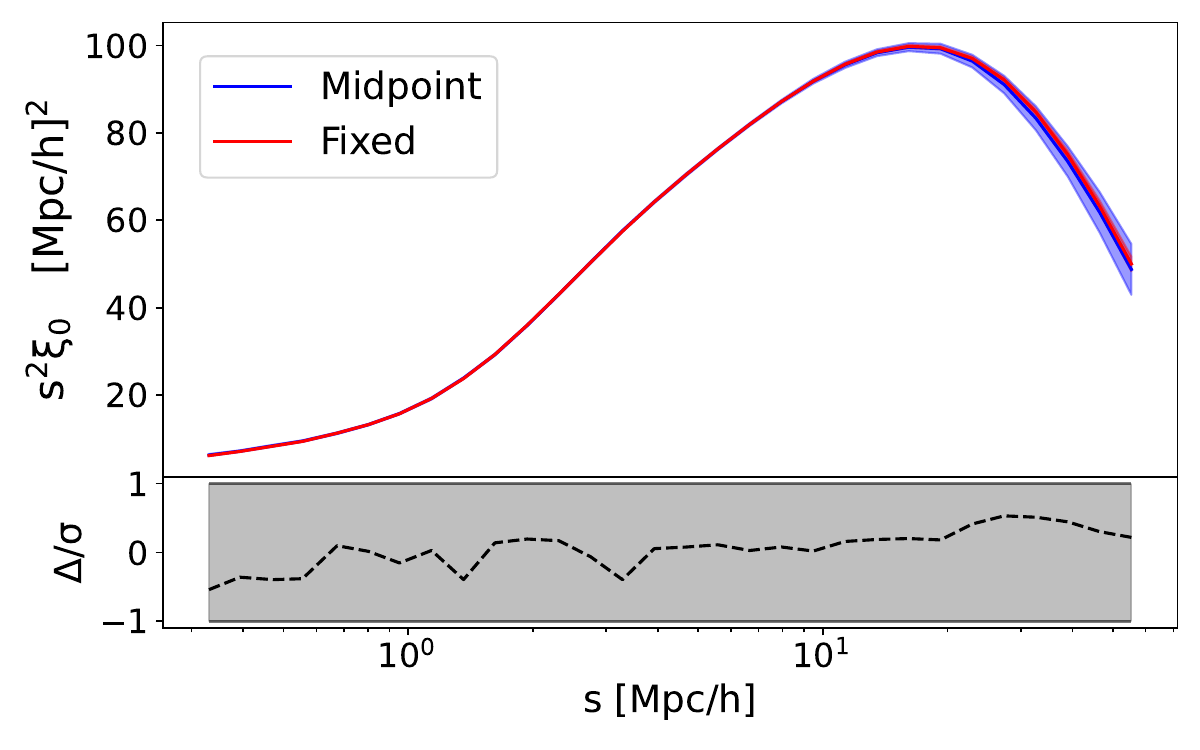}
	\caption{Mean clustering with its standard deviation of the Planck2018 25 boxes using the median HOD. In red $\xi_0$ is computed using Eq. \ref{eq:peebles} with a fixed LoS,and in blue using Eq. \ref{eq:LS} with a varying midpoint LoS. In the latter case in order to get a realistic clustering we place the observer at the center of the boxes, and cut a spherical full sky footprint from z = 0.15 to z=0.25. The bottom panel shows the residual using the cosmic variance of the (realistic) varying LoS.}
	\label{fig:wideangle}
\end{figure}

In figure \ref{fig:wideangle} we show the mean clustering with its standard deviation computed using Eq. \ref{eq:peebles} with a fixed LoS in red, and using Eq. \ref{eq:LS} with a varying midpoint LoS in blue. In the latter case, in order to get a realistic clustering we place the observer at the center of the box, and cut a spherical full sky footprint from $z = 0.15$ to $z=0.25$ in radial distance. The bottom panel shows the residual using the cosmic variance of the varying LoS which is of course larger an more realistic. The shaded area correspond to a $1\sigma$ deviation. We see that although the deviation seems to increase as the separation $s$ gets larger, the difference is well within the cosmic variance. For the considered redshift and separation bins, the Peebles estimator gives an unbiased description of the clustering.


\begin{figure}
	\centering
	\includegraphics[width=1.\columnwidth]{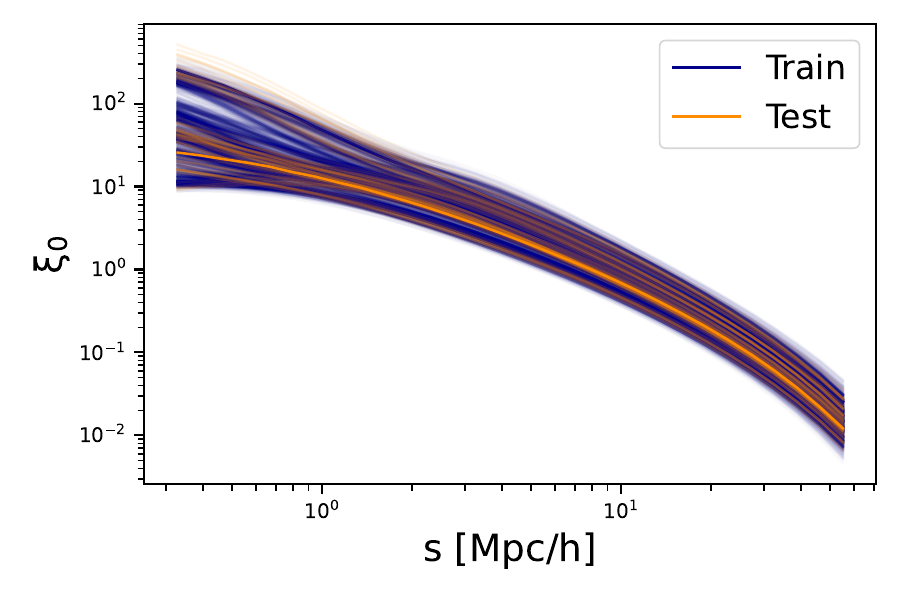}
	\caption{Emulator train and test sets in blue and orange respectively. The clustering for every cosmology and HOD is measured using Eq. \ref{eq:peebles} and applying RSD along a fixed LoS that we choose to be the z axis of the box.}
	\label{fig:gg_mono_dataset}
\end{figure}

Figure \ref{fig:gg_mono_dataset} shows the clustering measured for every cosmology and HOD using Eq. \ref{eq:peebles} and applying RSD along a fixed LoS that we choose to be the z axis of the box.
A sample of the train and test sets are shown in blue and orange respectively. We choose the separation range to be a logarithmic binning going from the resolution of the simulation $0.3~\hmpc$ to $60~\hmpc$ where the flat sky approximation is still holding. Hereafter we define as $X_\textrm{s}$ the vector containing those separations (used in section~\ref{sec:ML}). 
To reduce shot noise, we populate five times the halos using the same HOD model, only changing the random seed, and we average their clustering. 

\begin{figure}
	\centering
	\includegraphics[width=1.\columnwidth]{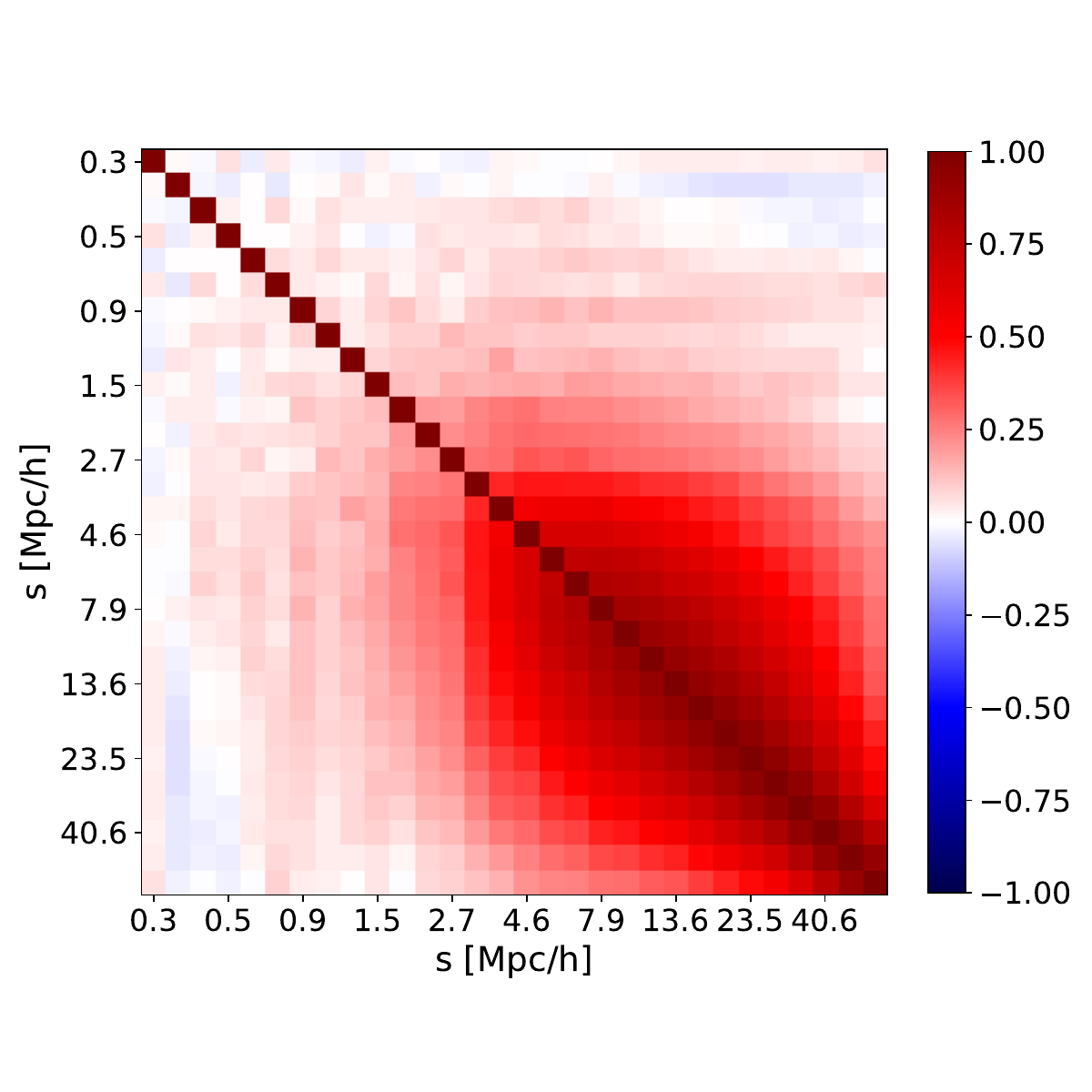}
	\caption{Correlation matrix of the 2PCF monopole computed from 1400 small boxes realisations of the Planck2018 cosmology, and using the median HOD. The covariance is rescaled to match the volume of the base boxes.}
	\label{fig:cov}
\end{figure}

Although \textsc{AbacusSummit} covers a large volume, the train and test cosmology boxes are run with the same initial phase, and the measured clustering is still subject to cosmic variance. To quantify the uncertainties on our measurements, we measure the 2PCF on 1400 of the small Planck2018 boxes using the median HOD and rescaling the volume, and build the covariance matrix $C_{{\rm cosmic}}$. The corresponding correlation matrix is shown in figure \ref{fig:cov}. We see that while the scales below a few $\hmpc$ are almost uncorrelated, the large scale correlations should not be neglected.

\section{Multi-scale Gaussian process Model}
\label{sec:ML} 


In the following sections we give a breve description of GP formalism, and introduce our new model allowing to consider every scales at the same time. For a more complete discussion on GP see \cite{rasmussen_gaussian_2006}.


\subsection{Standard GP regression model}
\label{sec:ML:IGP}

Let $\mathcal{D} = \{ (\vec{x_i}, y_i)|_{i=1}^n \} $ be a training set composed of $n$ pairs of $d$-dimension input parameter vectors $\vec{x_i}$ and scalar observations $y_i$.
Let $X$ be the $n \times d$ matrix concatenating every input vector and $Y$ a $n$ column vector. We suppose that the observed values $y_i$ are given by 
\begin{equation}
y_i = f(\vec{x_i}) + \epsilon
\label{eq:gp_1}
\end{equation}

\noindent where $f$ is the true underlying mapping function we wish 
to learn and $\epsilon$ is an additive noise following an independent 
identically distributed Gaussian distribution with mean zero and variance $\sigma_n^2$.

\begin{equation}
\epsilon \sim \mathcal{N} \left ( 0, \sigma_n^2 I\right ).
\label{eq:gp_2}
\end{equation}
where $I$ is the identity matrix. We assume that the mapping function can be described as 

\begin{equation}
f(\vec{x}) = \Phi(\vec{x})^\top \vec{w}
\label{eq:gp_22}
\end{equation}

\noindent where $\Phi(\vec{x})$ is a set of $n_f$ feature functions  (e.g. polynomials $\Phi(\Vec{x}) = (||\Vec{x}||, \ ||\Vec{x}||^2, \ ||\Vec{x}||^3, \ ..., \ ||\Vec{x}||^{n_f})$) mapping 
the $d$-dimensional input vector into a $n_f$-dimensional feature space, 
and $\vec{w}$ is a $n_f$-dimension vector of unknown weights (or parameters).
This model is a linear function of $\vec{w}$ for any set of feature functions
hence easier to handle analytically in the training process for instance.

The likelihood of observing $Y$ 
for a set of input parameters $X$ and weights $\vec{w}$ can then be written as following a Gaussian distribution
\begin{equation}
p \left ( Y | X, \vec{w}\right ) \sim \mathcal{N} \left (\Phi(X)^\top \vec{w}, \sigma_n^2 I  \right )
\label{eq:gp_3}
\end{equation}
Choosing a Gaussian prior with zero mean and covariance $\Sigma$ for the weights 
\begin{equation}
p \left ( \vec{w} \right ) = \mathcal{N}(0,\Sigma)
\label{eq:gp_4}
\end{equation}
and following Bayes' theorem we can write the posterior distribution 
over the weights as
\begin{equation}
p \left ( \vec{w}|Y,X \right ) = \frac{p \left ( Y|X,\vec{w} \right ) p \left ( \vec{w} \right )}{p \left ( Y|X \right )}
\label{eq:gp_5}
\end{equation}
where $p \left ( Y|X \right )$ is the evidence or marginal likelihood. 

Making $n_*$ predictions $f_*$ for a new set of input vector $X_*$ (with $X_*$ a $n_* \times d$ matrix) requires 
to marginalise over the weight distributions. 
The evidence is a just normalisation constant. Using the Gaussian properties of the prior and likelihood, an analytical expression for the posterior distribution of the prediction can be derived. It is expressed as:
\begin{equation}
\begin{aligned}
p \left ( f_*|\vec{x_*}, X, Y \right ) &= \int p \left ( f_*|\vec{x_*}, \vec{w} \right ) p \left ( \vec{w} |X,Y \right ) \mathrm{d} \vec{w} \\
&  \sim \mathcal{N} \left( \mu_f , \Sigma_f \right)
\end{aligned}
\label{eq:gp_6}
\end{equation}
where 
\begin{eqnarray}
    \mu_f &= & \Phi_*^\top \Sigma \Phi \left(\Phi^\top \Sigma \Phi + \sigma_n^2 I\right)^{-1} Y, \\
    \Sigma_f & = &  \Phi_*^\top \Sigma \Phi_* - \Phi_*^\top \Sigma \Phi \left(\Phi^\top \Sigma \Phi + \sigma_n^2 I\right)^{-1} \Phi^\top \Sigma \Phi_* 
\end{eqnarray}
and $\Phi = \Phi(X)$ and $\Phi_* = \Phi(X_*)$. 
This posterior is a normal probability distribution over a function space.
Note that making a prediction at some new location of the input space is
equivalent to analytically performing a Bayesian inference. 
Thanks to the linearity of the model and the Gaussian assumptions,
the posterior probability can be analytically computed.


Let us now define the kernel $k(\vec{x},\vec{x'}) = \Phi(\vec{x})^\top \Sigma \Phi(\vec{x'})$ (Note that $k(\vec{x},\vec{x'})$ returns a scalar). Doing so allows us to switch from the weight-space view to the so called function-space view. The expressions for the mean and covariance of the model posterior become:



\begin{equation}
\begin{aligned}
\mu_f  &= K(X_*,X) \left ( K(X,X) + \sigma_n^2 I\right )^{-1} Y ,\\
\Sigma_f &= K(X_*,X_*) - K(X_*,X) \left ( K(X,X) + \sigma_n^2 I\right )^{-1}  K(X,X_*)
\end{aligned}
\label{eq:gp_8}
\end{equation}
where $K(X,X_*)$ denotes the $n \times n_*$ covariance matrix built by evaluating the kernel $k(\vec{x}, \vec{x_*})$ for the different values of $\vec{x}$ and $\vec{x'}$ (i.e $K_{\rm ij} = k(\vec{x}_i, \vec{x_*}_j)$).  

Specifying a family of feature functions, and letting their number $n_f$ go to infinity leads to compact form of kernels. The more popular in the literature (obtained using Gaussian feature functions) is the squared exponential kernel 
\begin{equation}
k_{\mathrm{exp}}(r) = \exp \left ( \frac{-r^2}{2}\right )
\label{eq:gp_exp}
\end{equation}
or the Matern kernels
\begin{equation}
\begin{aligned}
&k_{\rm{M_{3/2}}}(r) = \left(1 + \sqrt{3}r  \right )\rm{exp} \left ( -\sqrt{3}r\right ) \\
&k_{\rm{M_{5/2}}}(r) = \left(1 + \sqrt{5}r + \frac{5}{3}r^2  \right )\rm{exp} \left ( -\sqrt{5}r\right ) .
\end{aligned}
\label{eq:gp_matern}
\end{equation}

These are stationary kernels, functions of the distance between two points in the input space, commonly defined as $r(\vec{x},\vec{x'}) = (\vec{x} - \vec{x'})^\top M (\vec{x} - \vec{x'})$ with $M = \vec{l}^{-2} I$, where $\vec{l}$ is a  $d$-dimensional vector containing characteristic length-scales across each dimension of the input space. 
A large value for some component of $\vec{l}$ 
(compared to the input space variance) describes a smoothly varying signal in the corresponding dimension, thus allowing for a larger range on interpolation. In practice, every kernel is also scaled by an additional hyperparameter $\sigma^2$ to account for the signal amplitude in the output space. Note that the sum or product of different kernels also give a valid kernel. The training of the GP consists in choosing one or more kernels and inferring the hyperparameters $\vec{\theta} = \{\vec{l}, \sigma \}$ from the input parameter matrix $\vec{X}$ and their corresponding output observations $Y$. This is equivalent to a two level hierarchical Bayesian inference where Eq \ref{eq:gp_5} is rewritten 
\begin{equation}
p \left ( \vec{w}|Y,X,\vec{\theta} \right ) = \frac{p \left ( Y|X,\vec{w} \right ) p \left ( \vec{w}|\vec{\theta} \right )}{p \left ( Y|X,\vec{\theta} \right )}.
\label{eq:gp_9}
\end{equation}

Now maximising the posterior probability for new predictions requires to maximize the evidence $p \left ( Y|X, \boldsymbol{\theta} \right )$. This term now acts as a likelihood in the second level of the hierarchical Bayesian inference. The posterior over the hyperparameters can then be expressed as

\begin{equation}
p \left (\vec{\theta}  |  Y, X \right ) = \frac{p \left( Y  | X, \vec{\theta}  \right) p( \vec{\theta})}{p \left( Y |  X \right)}
\label{eq:gp_2nd_post}
\end{equation}
where $p \left ( \vec{\theta}  \right )$ is the hyper-prior and $p \left (Y  |  X \right)$ is the hyper-evidence. This second level inference is usually not analytically tractable and requires algorithmic maximisation of the log-likelihood

\begin{equation}
\begin{aligned}
\log \left[  p ( Y|X, \vec{\theta} ) \right] = &-\frac{1}{2} Y^\top \left ( K(\vec{x},\vec{x}) + \sigma_n^2 I\right )^{-1}Y \\& - \frac{1}{2} \log \left|K(\vec{x},\vec{x}) + \sigma_n^2 I \right| - \frac{n \log (2\pi)}{2}.
\end{aligned}
\label{eq:gp_logL}
\end{equation}

The first term in the right-hand side of the equation above describes the goodness of the fit while the second term 
is the complexity penalty which is independent from the observables $Y$, and allows (partially) GP models to avoid overfitting. 
We recall the standard result for the gradient of the log-likelihood with respect to a given hyperparameter $\theta_i$: 

\begin{equation}
\begin{aligned}
\frac{\partial \log p}{\partial \theta_i} = \frac{1}{2} \left( \vec{\alpha}^\top \frac{\partial K_y}{\partial \theta_i}\vec{\alpha} - \textrm{Tr} \left( K_y^{-1} \frac{\partial K_y^\top}{\partial \theta_i} \right) \right)
\end{aligned}
\label{eq:gp_grad_logL}
\end{equation}
with $K_y = \left(K + \sigma_n^2 I \right)$ and $\vec{\alpha} = K_y^{-1}Y$.


Training a GP (learning the hyperparameters) requires to compute this likelihood and its gradient several times, which involve getting the inverse and the determinant of the $n \times n$ matrix $K_y$. 
For a training set larger than a few thousands of samples, these operations can become computationally prohibitive.  
For example, considering our case described in section~\ref{sec:data}, this matrix
would have $n = n_\Omega \times n_\mathrm{HOD} \times n_s = 88 \times 600 \times 30 \sim 10^6$, where $n_\Omega$, $n_\mathrm{HOD}$ and $n_s$ are the number of sampled cosmologies, HODs and separations, respectively.

\subsection{Multi-scale GP model}
\label{sec:ML:MGP} 

To overcome the issue of large matrix inversions for the standard GP, we can use a particular way of sampling the training data input parameters. Those are known as Kronecker GP models, and we describe them in the following.

Let the input parameter matrix of the training set $X$ be  composed of $n$ samples of $d$ dimensions. If this matrix  can be decomposed as a Kronecker product $X=X_1 \otimes X_2$ where $X_1$ and $X_2$ have smaller dimensions ($n_1\times d_1$ and $n_2 \times d_2$ respectively) with $n = n_1 \times n_2$ and $d = d_1 + d_2$. This decomposition is only possible if the training set is grid-sampled (eg. the same HOD parameters are repeated for each set of cosmological parameters), which is sub-optimal, 
but as a result the covariance kernel can be computed as 
a Kronecker product of two lower dimension matrices 
$K = K_1 \otimes K_2$.

This structure allows to make use of the algebraic properties of Kronecker products. If $K$ is positive definite so it can be written as a principal component decomposition $K = Q\Lambda Q^\top$, where $Q = Q_1 \otimes Q_2$ and $\Lambda = \Lambda_1 \otimes \Lambda_2$, with $K_i = Q_i\Lambda_i Q_i^\top$, with every  $\Lambda$ matrix diagonal.
Thus every calculation can be performed in sub dimensional spaces.
Such a model was implemented in the Python library \textsc{GPy} \footnote{\url{https://github.com/SheffieldML/GPy}} but only for Kronecker decompositions into two sub-spaces.
\cite{saatci_scalable_nodate} gives a generalisation of this framework to $k$ decompositions $X = \bigotimes_{i=1}^{k}X_i$ but valid only for a diagonal uncorrelated noise matrix $N = \sigma^2 I$.

In this work, we extended this framework to general correlated noise covariance matrices of the form $N =\bigotimes_{i=1}^{k}N_i$. 
This is important in clustering modelling because the noise covariance matrices are generally far from being diagonal as seen in figure \ref{fig:cov}.
The essential step is now to compute the inverse and the determinant
of $\left ( K + N \right )$ in Eq \ref{eq:gp_logL}. 
Assuming a Kronecker structured training set and performing the decompositions $N_i = U_i S_i U_i^\top$ we can derive

\begin{equation}
\begin{aligned}
&\left ( K + N \right )^{-1} =\left ( \bigotimes_{i=1}^{k}K_i + \bigotimes_{i=1}^{k}N_i \right )^{-1} \\
&= \left ( \bigotimes_{i=1}^{k}K_i + \bigotimes_{i=1}^{k}U_iS_iU_i^\top \right )^{-1} \\
&= \left ( \bigotimes_{i=1}^{k}K_i + \bigotimes_{i=1}^{k}U_i \bigotimes_{i=1}^{k}S_i \bigotimes_{i=1}^{k}U_i ^\top \right )^{-1} \\
&=\bigotimes_{i=1}^{k}U_i \bigotimes_{i=1}^{k}S_i^{-\frac{1}{2}}  \left ( \bigotimes_{i=1}^{k}\Tilde{K_i} + \bigotimes_{i=1}^{k} I_i \right )^{-1} \bigotimes_{i=1}^{k}S_i^{-\frac{1}{2}} \bigotimes_{i=1}^{k}U_i^\top \\
&=\bigotimes_{i=1}^{k}U_i \bigotimes_{i=1}^{k}S_i^{-\frac{1}{2}}\bigotimes_{i=1}^{k}Q_i \left ( \bigotimes_{i=1}^{k}\Lambda_i + \bigotimes_{i=1}^{k} I_i \right )^{-1} \\
& \hspace{5cm}\bigotimes_{i=1}^{k}Q_i^\top\bigotimes_{i=1}^{k}S_i^{-\frac{1}{2}} \bigotimes_{i=1}^{k}U_i^\top 
\end{aligned}
\label{eq:gp_log_mkgp}
\end{equation}
where we used different properties of the Kronecker product, and defined the projected matrices $\Tilde{K_i} = S_i^{-\frac{1}{2}} U_i^\top K_i U_i S_i^{-\frac{1}{2}}$ that once built can be decomposed as $\Tilde{K_i} = Q_i \Lambda_i Q_i^\top$. 



Now $\Lambda = \bigotimes_{i=1}^{k}\Lambda_i$ is trivial to build as $\Lambda_i$ are all diagonal matrices. Moreover $\bigotimes_{i=1}^{k} I_i = I$, thus the $n \times n$ diagonal matrix $\left (\Lambda + I \right )$ is easily built and inverted. 
Additionally $\left | \bigotimes_{i=1}^{k} U_i \right | = 1 $ and $ \left | \bigotimes_{i=1}^{k} Q_i \right | = 1$, so the determinant from Eq \ref{eq:gp_logL} is simply the product of the eigenvalues products $\left | S^{-1/2} \right| \left |\Lambda + I \right| \left | S^{-1/2} \right|$. The goodness-of-fit term is computed by iteratively using the \textsc{kronmvprod} algorithm (described in \cite{saatci_scalable_nodate}) that effectively evaluates operation of the form $\vec{\alpha} = \left (\bigotimes_{i=1}^{k}A_i \right)\vec{b}$ with $\vec{b}$ a column vector.

We also require the evaluation of log-likelihood gradients. The first term of Eq \ref{eq:gp_grad_logL} can be easily computed as the hyperparameters are specific to a particular sub-dimension $d$, hence to a kernel $K_d$, i.e.,
\begin{equation}
\begin{aligned}
&\frac{\partial K}{\partial \theta_i} = \frac{\partial K_i}{\partial \theta_i}\otimes \left( \bigotimes_{j \neq i} K_j \right )\\
&\frac{\partial N}{\partial \theta_i} = \frac{\partial N_i}{\partial \theta_i}\otimes \left( \bigotimes_{j \neq i}N_j \right ).
\end{aligned}
\label{eq:gp_grad1}
\end{equation}
while the second term of Eq \ref{eq:gp_grad_logL} can be computed using the cyclic properties of the trace operator:
\begin{equation}
\begin{aligned}
&\textrm{Tr} \left( K^{-1} \frac{\partial K^\top}{\partial \theta_i} \right) = \textrm{diag}\left ( \bigotimes_{i=1}^{k}\Lambda_i + \bigotimes_{i=1}^{k} I_i \right )^\top \textrm{diag}\left(H^\top \frac{\partial K^\top}{\partial \theta_i} H \right)\\
&\textrm{Tr} \left( N^{-1} \frac{\partial N^\top}{\partial \theta_i} \right) = \textrm{diag}\left ( \bigotimes_{i=1}^{k}\Lambda_i + \bigotimes_{i=1}^{k} I_i \right )^\top \textrm{diag}\left(H^\top \frac{\partial N^\top}{\partial \theta_i} H \right)
\end{aligned}
\label{eq:gp_grad2}
\end{equation}
with $H = \bigotimes_{i=1}^{k}U_i \bigotimes_{i=1}^{k}S_i^{-\frac{1}{2}}\bigotimes_{i=1}^{k}Q_i $. 
The last terms in above equations are given by 
\begin{equation}
\begin{aligned}
&H^\top \frac{\partial K^\top}{\partial \theta_i} H = H_i^\top \frac{\partial K_i^\top}{\partial \theta_i} H_i \otimes \left( \bigotimes_{i\neq j} H_j^\top \frac{\partial K_j^\top}{\partial \theta_i} H_j \right)\\
&H^\top \frac{\partial N^\top}{\partial \theta_i} H = H_i^\top \frac{\partial N_i^\top}{\partial \theta_i} H_i \otimes \left( \bigotimes_{i\neq j} H_j^\top \frac{\partial N_j^\top}{\partial \theta_i} H_j \right) 
\end{aligned}
\label{eq:gp_grad3}
\end{equation}

We provide an implementation of this model using \textsc{GPy} framework and based on \textsc{JAX} GPU parallelisation for algebra operations. It is publicly available as the \textsc{MKGpy}\footnote{\url{https://gitwhub.com/TyannDB/mkgpy/tree/main}} package. 

\subsection{Application to galaxy clustering}
\label{sec:ML:app}

To apply this model to our observable, we decompose the parameter space as a gridded structure $X = X_{\Omega} \otimes X_\textrm{HOD} \otimes X_\textrm{s}$.
Specifically, as mentioned in section \ref{sec:data:hod}, we measure the correlation function on the same scales using the same HOD sampling for every cosmology, allowing us to decompose the signal and noise covariances as 

\begin{equation}
\begin{aligned}
&K = K_{\Omega} \otimes K_\textrm{HOD} \otimes K_\textrm{s} \\
&N = N_{\Omega} \otimes N_\textrm{HOD} \otimes N_\textrm{s}.
\end{aligned}
\label{eq:gp_emu_kernels}
\end{equation}

We found that the optimal signal kernels (maximizing the log-likelihood) are simply two Matern$_{3/2}$ kernels for cosmological and HOD spaces and a squared exponential kernel for the scales space.
This  choice of modeling leads to 16 signal hyperparameters $\theta = \left\{ \Vec{l_\Omega},\Vec{l_\textrm{HOD}},l_\textrm{s},\sigma \right\} $ composed of 
9 lengthscales for the cosmological parameters $\vec{l}_\Omega$, 5 for the HOD parameters $\vec{l}_\mathrm{HOD}$, one for the separations $l_s$ and one overall variance of the signal $\sigma$.
For the noise kernels we chose $N_\textrm{s}$ to be the fixed covariance matrix of the 2PCF monopole computed from the 1400 small boxes (figure \ref{fig:cov}).
For the cosmological and HOD noise kernels with chose diagonal heteroscedastic kernels of the form $N(x,x') = \Vec{\sigma}_n(x)^2 \delta_{\textrm{D}}(x-x') I$, allowing a different variance for every HOD and cosmological configuration. 
We let $\Vec{\sigma_n}$ to be fully parameterized vectors, leading to $88$ noise hyperparameters for each cosmology and $600$ for each HOD. Those noise hyperparameters $\left ( \Vec{\sigma_{n_\Omega}}, \Vec{\sigma_{n_\textrm{HOD}}} \right ) $ are fitted along with the signal hyperparameters $\theta$.

To avoid mathematical overflow and ease the hyperparameter learning, 
we normalise each dimension $i$ of the input and output training sets as follow

\begin{equation}
\begin{aligned}
&\tilde{X}_{\Omega}^i = \frac{X_{\Omega}^i - \textrm{min}(X_{\Omega}^i)}{ \textrm{max}(X_{\Omega}^i) -  \textrm{min}(X_{\Omega}^i)}\\
&\Tilde{X}_\textrm{HOD}^i = \frac{X_{\textrm{HOD}}^i - \textrm{min}(X_{\textrm{HOD}}^i)}{ \textrm{max}(X_{\textrm{HOD}}^i) -  \textrm{min}(X_{\textrm{HOD}}^i)}\\
&\Tilde{X}_\textrm{s} = \frac{\textrm{log}(X_{\textrm{s}}) - \textrm{log}(\textrm{min}(X_{\textrm{s}}))}{ \textrm{log}(\textrm{max}(X_{\textrm{s}})) -  \textrm{log}(\textrm{min}(X_{\textrm{s}}))}
\end{aligned}
\label{eq:gp_xnorm}
\end{equation}

$X_{\Omega}$ and $X_{\rm HOD}$ are normalized to be distributed in the range $\left [0 , 1 \right ]$.
Because we use stationnary kernels $k(x,x') = k(|x-x'|)$ we need to make sure that the variance of the signal is roughly constant as a function of the input $x$, which is not the case along the subspace $X_{\rm s}$. 
For this input dimension we use a logarithmic transformation to flatten the signal before applying the normalisation. 
We also normalise the output training set $Y$ 
to be normally distributed around zero with a variance of one.
Because of the large range of magnitude span by the 2PCF, we also
apply a logarithmic transformation before doing so. 
\begin{equation}
\Tilde{Y} = \frac{\textrm{log}(Y) - \textrm{mean}(\textrm{log}(Y))}{ \textrm{std}(\textrm{log}(Y))}
\label{eq:gp_ynorm}
\end{equation}
The covariance matrix used as $N_\textrm{s}$ must also be projected in the normalised space. To do so we use the error propagation 
\begin{equation}
\Tilde{N}_\textrm{s} = J N_\textrm{s} J^\top
\label{eq:gp_jacobian}
\end{equation}
with $J$ the mapping Jacobi matrix $J_{ij} = \frac{\partial \Tilde{Y}_i}{\partial Y_j}$ computed for the mean of the 1400 small boxes. 
Once the emulator is trained, every new input configuration $X_\textrm{test}$ should be normalised using Eqs \ref{eq:gp_xnorm}, and every model prediction $\mu_f$ and its covariance $\Sigma_f$ (Eq. \ref{eq:gp_8}) should be transformed back into the real space using Eqs \ref{eq:gp_ynorm} and \ref{eq:gp_jacobian}. 

The hyperparameters are optimised using the \textsc{Scipy} gradient based algorithm \textsc{lbfgs} with 10 random initialisations and selecting the best-fit hyperparameters yielding the highest likelihood. 
To compare performances, we train both the standard GP where each separation is treated independently (as separate emulators) and our new multi-scale GP. We employ the same kernels, normalisation and optimization method.


\section{Emulator performance}
\label{sec:emu_perf} 

In this section we compare the performances of two emulators. First, 
the independent Gaussian process model (IGP), where each scale $s$ of the correlation 
function $\xi(s)$ has its own independent GP depending on cosmological and HOD 
parameters. And second, our multi-scale Gaussian process model (MGP) able to predict the full model for $\xi(s)$ as well as an associated non-diagonal model covariance.

\subsection{Prediction accuracy}
\label{sec:emu_perf:test} 

We computed predictions for the test-set composed of 20 HOD $\times$ 6 
cosmologies (see section \ref{sec:data} for details) using both IGP and MGP models, and compared with their observed correlation function monopoles.

\begin{figure}
	\centering
	\includegraphics[width=1.\columnwidth]{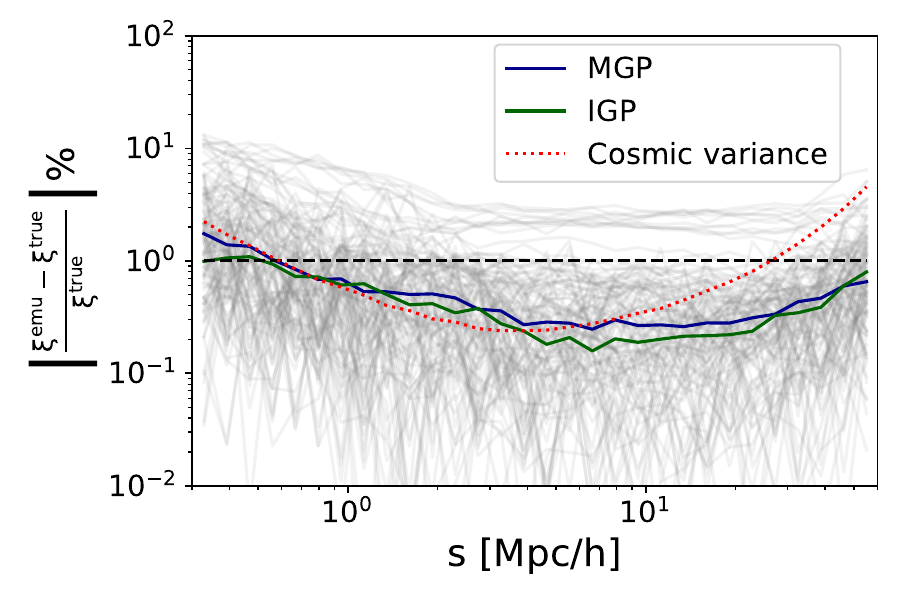}
	\caption{Grey lines show the relative difference in $\%$ between the MGP emulator prediction and the expected values for the full test set. The blue and green lines are the median of those deviations for the MGP and IGP models respectively. The black dotted line correspond to 1$\%$ deviation. The red dotted line shows the median of $| \sigma/\xi^{true} |$ in \%. With $\sigma^2$ the cosmic variance computed for the Planck 2018 cosmology and median HOD.}
	\label{fig:model_perf_full}
\end{figure}

Figure \ref{fig:model_perf_full} describes the IGP and MGP performances using the full test set. The grey lines show the relative difference in $\%$ between the MGP emulator prediction and the expected values from the test set. 
The black dotted line correspond to a 1$\%$ deviation. 
The blue and green lines are the median of those deviations for the MGP and IGP models respectively, assessing a subpercent precision for both models on most of the scales. 
However, the correlation functions of our test set are noisy due to cosmic variance and shot noise, so we do not expect the difference $\Delta \equiv \xi^\mathrm{emu}-\xi^\mathrm{true} $ to be smaller than the estimated uncertainty of $\xi^\mathrm{true}$ . 
The red dashed line shows the median of $\rm \left |\frac{\sigma}{\xi^{true}} \right |$ in $\%$ with $\sigma$ the diagonal of $C_\textrm{cosmic}$ . 
A difference smaller than this reference would indicate overfitting of the particular initial phase of the simulations, while larger differences would show that the prediction performance is poor. Both MGP and IGP models seems to be slightly overfitting the initial phase for the large scales, but follow the expected accuracy on small scales. 
Note that for the very small scales, the MGP is less overfitting the test set than the IGP.
We see that the relative error of the full sample in grey varies around the cosmic variance expectation. 
 Note that the dotted red line corresponds to the cosmic variance of the central cosmology and HOD. The wide HOD scatter causes large variations to the clustering amplitude and hence to the cosmic variance.

\subsection{Robustness to cosmic variance}
\label{sec:emu_perf:phase} 

\begin{figure}
	\centering
	\includegraphics[width=1.\columnwidth]{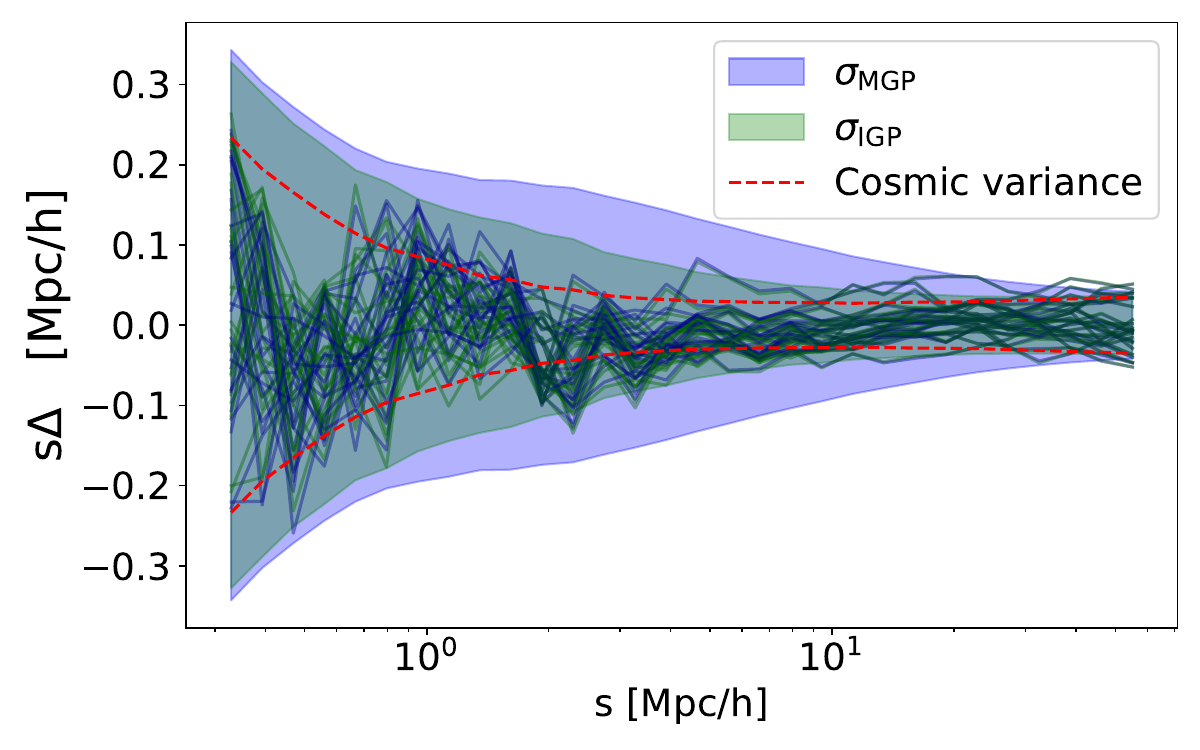}
	\caption{Difference between the emulators prediction and the 25 realisations of Planck2018 cosmology with the median HOD in blue and green lines for the MGP and IGP respectively. The red dashed line describes the cosmic variance level. The total variances including the emulators variance predictions are shown with the blue and green shaded area.}
	\label{fig:error_pred_25boxes}
\end{figure}

Our models have been trained on correlation functions from simulations ran 
with different cosmologies and HOD parameters, though all started from  
the same initial phase. 
Over-fitting this particular phase could potentially bias our clustering 
predictions on simulations with different initial phases.
We tested this issue with 25 phase realisations of AbacusSummit ran with the 
same Planck2018 cosmology and the same set of HOD parameters.

Gaussian processes provide not only a prediction for a model, but also 
an estimate for the model uncertainty that depends on the input parameters $\vec{\theta}$. This uncertainty is expressed as a covariance matrix, noted $C_{{\rm emu}}(\vec{\theta})$.
When performing inferences on real data, this model covariance has to be added to the 
measurement uncertainty for the considered scales $C_\textrm{cosmic}$ as:
\begin{equation}
C_{{\rm tot}}(\rm \theta_{\Omega},\theta_{hod}) = C_{{\rm cosmic}} + C_{{\rm emu}}(\rm \theta_{\Omega},\theta_{hod}).
\label{eq:cov_tot}
\end{equation}
We deliberately dropped the dependency of $C_{{\rm cosmic}}$ on $\vec{\theta}$ as it is often considered fixed during inferences. 

Figure \ref{fig:error_pred_25boxes} shows the difference $\Delta$ between  the 25 measurements and predictions from both the MGP (blue) and the IGP (green) emulators. 
The red dashed line displays the cosmic variance level while blue and green lines show the total variance (model + cosmic) according to Eq.~\ref{eq:cov_tot}. 
We see that for intermediate scales, $\Delta$ is larger than the cosmic variance level, while using $C_{\rm tot}$ lowers the deviation to less than one sigma.
We also note that for this example the variance of the MGP is larger than the one from the IGP, however figure~\ref{fig:error_pred_25boxes} only presents the diagonal elements of the covariance matrices.

\begin{figure}
	\centering
	\includegraphics[width=1.\columnwidth]{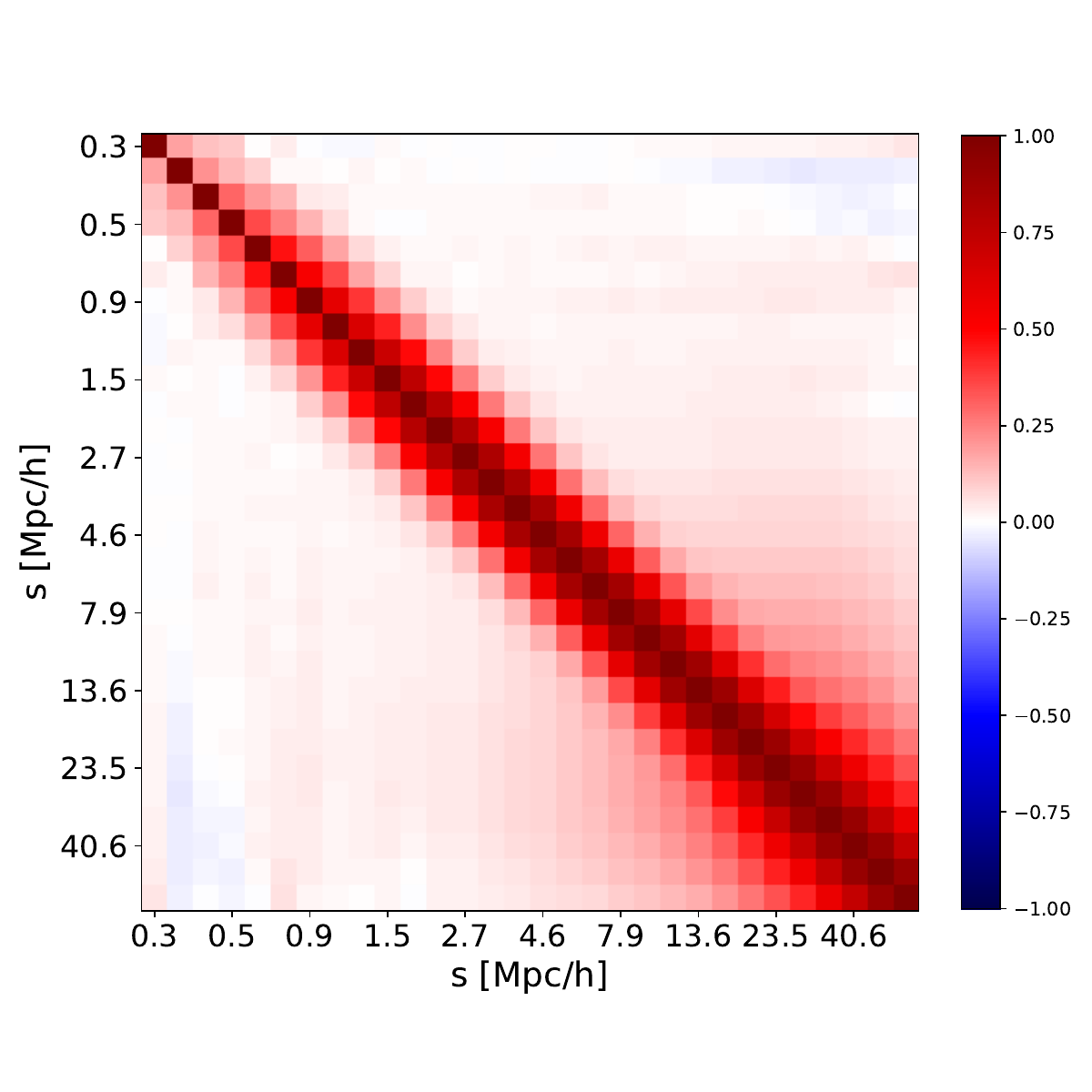}
	\caption{Total correlation matrix of the MGP including the model uncertainty prediction $C_{{\rm tot}}  = C_{{\rm cosmic}} + C_{{\rm emu}}(\rm \theta_{\Omega},\theta_{hod})$. The measurement uncertainty $C_{{\rm cosmic}}$ was evaluated for the Planck2018 cosmology with the median HOD using the 1400 small boxes. Including $C_{{\rm emu}}(\rm \theta_{\Omega},\theta_{hod})$ significantly correlates the small scales.}
	\label{fig:gg_mono_covtot}
\end{figure}

Figure~\ref{fig:gg_mono_covtot} shows the total (model + cosmic) correlation 
matrix of the MGP. 
Notice that compared to $C_{\rm cosmic}$ alone (Figure~\ref{fig:cov}), the small scales are now significantly correlated. The IGP's predicted covariance matrix is diagonal by construction so it cannot account for such correlations, potentially affecting cosmological constraints. We investigate this issue in section~\ref{sec:reco}. 

\begin{figure}
	\centering
	\includegraphics[width=1.\columnwidth]{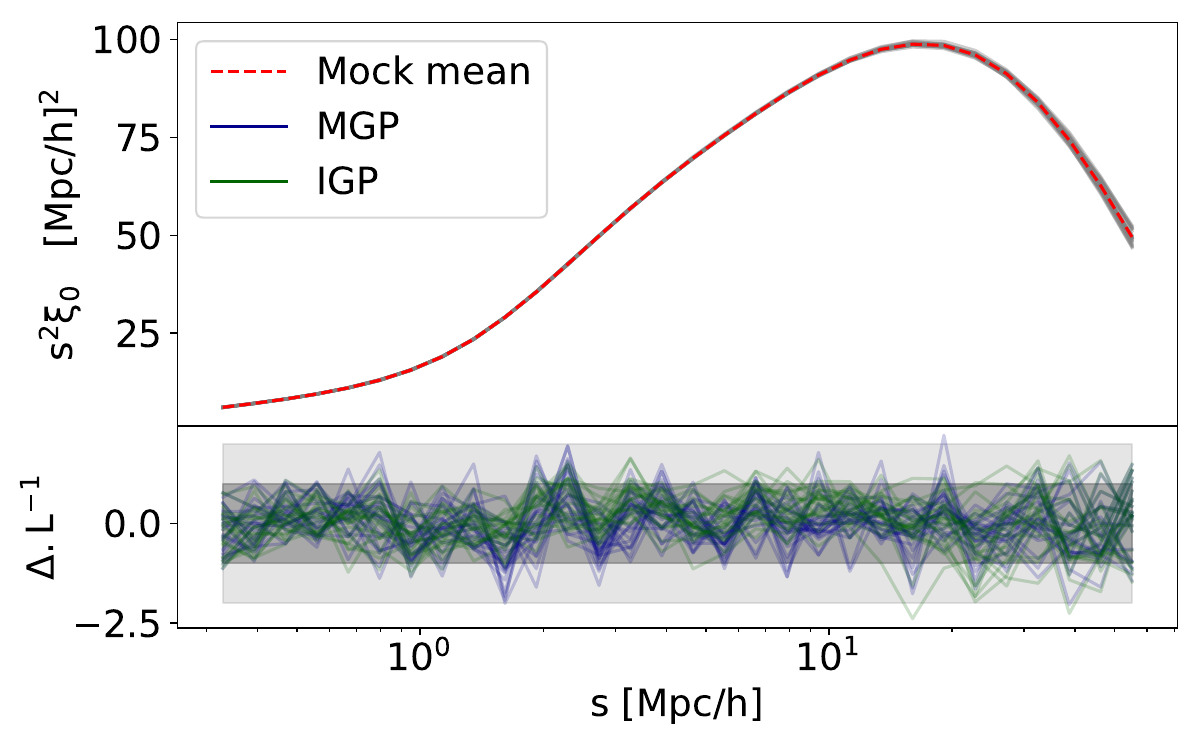}
	\caption{In the upper panel the 25 realisations of Planck2018 cosmology using the median HOD are shown in grey. The average is shown in red dotted line. The 25 residuals for both IGP (green) and MGP (blue) predictions are shown in the bottom panel. $L^{-1}$ is the Cholesky decomposition of the inverse of $C_\textrm{tot}$. The grey shaded areas correspond to one and two sigma deviations. }
	\label{fig:emu_accuracy}
\end{figure}

The 25 mocks clustering measurements are shown in the top panel of figure \ref{fig:emu_accuracy} with the average in red dotted line.
The bottom panel describes the 25 residual for both IGP (green) and MGP (blue) predictions. With $\Delta = \rm \xi^{emu} - \xi^{true}$ and $L^{-1}$ the Cholesky decomposition of the inverse of the total covariance $C_\textrm{tot}$. This residual metric is usefull as it takes into account the correlations between scales. The shaded area correspond to 1$\sigma$ and 2$\sigma$ deviation. For most of the scales, both MGP and IGP models are able to reproduce the clustering for the different realisations within  1$\sigma$ deviation. 
For every scales, the residuals are within 2$\sigma$ deviation. 
It seems that for both models the predicted uncertainty $C_\textrm{emu}$ solves the initial condition phase problem. 
We check in the next section the impact on the recovered cosmological and HOD parameters.

\section{Cosmology recovery and constraints}
\label{sec:reco} 

In this section we describe the performances of our models to recovering unbiased cosmological and HOD parameters
. 
We used the clustering measured on 25 realisations of the Planck2018 cosmology using the same median HOD parameters.

\subsection{Parameter inference using emulator model}
\label{sec:reco:like} 

The parameter inferences are performed by running MCMC chains with the library \textsc{emcee}, using a flat prior corresponding to Table~\ref{tab:cosmo_hod_range} and a standard Gaussian log-likelihood defined as

\begin{equation}
\log \mathcal{L}(\vec{\theta}) \propto -\frac{1}{2} \left(\vec{m(\theta) - \mu} \right) C^{-1}_{{\rm tot}}(\Vec{\theta}) \left(\Vec{m(\theta) - \mu} \right)^\top -\frac{1}{2} \textrm{log} |C_{{\rm tot}}(\Vec{\theta})|
\label{eq:likelihood}
\end{equation}
where $\vec{\mu}$ is the data vector, $\vec{m(\theta)}$ is the model prediction for a 
given set of parameters $\vec{\theta} = \left( \rm \Vec{\theta}_{\Omega},\Vec{\theta}_{hod} \right )$, and $C_{{\rm tot}}(\Vec{\theta})$ is the total (model + cosmic ) covariance defined in Eq.~\ref{eq:cov_tot}.
Compared to an usual cosmological inference, the terms $C^{-1}_{{\rm tot}}(\Vec{\theta})$ and $\textrm{log} |C_{{\rm tot}}(\Vec{\theta})|$ must be computed for every iteration because of the emulator covariance dependence on the parameters. 
To do so efficiently, we make use of the fast and stable Cholesky decomposition.


For a given set of cosmological parameters we can infer the corresponding growth rate $f$.
The uncertainty on the measured growth rate is simply propagated as $\sigma_f = \frac{\partial f}{\partial \Omega_i} C_{{\Omega}_{ij}} \frac{\partial f}{\partial \Omega_j}$, where $C_{{\Omega}_{ij}}$ is the covariance of the inferred cosmological parameters. 
We use the \textsc{jacobi}\footnote{\url{https://hdembinski.github.io/jacobi/}} python library to compute our model's gradient.

\subsection{Cubic boxes : Average fit}
\label{sec:reco:average}

\begin{figure*}
	\centering
	\includegraphics[width=1\textwidth]{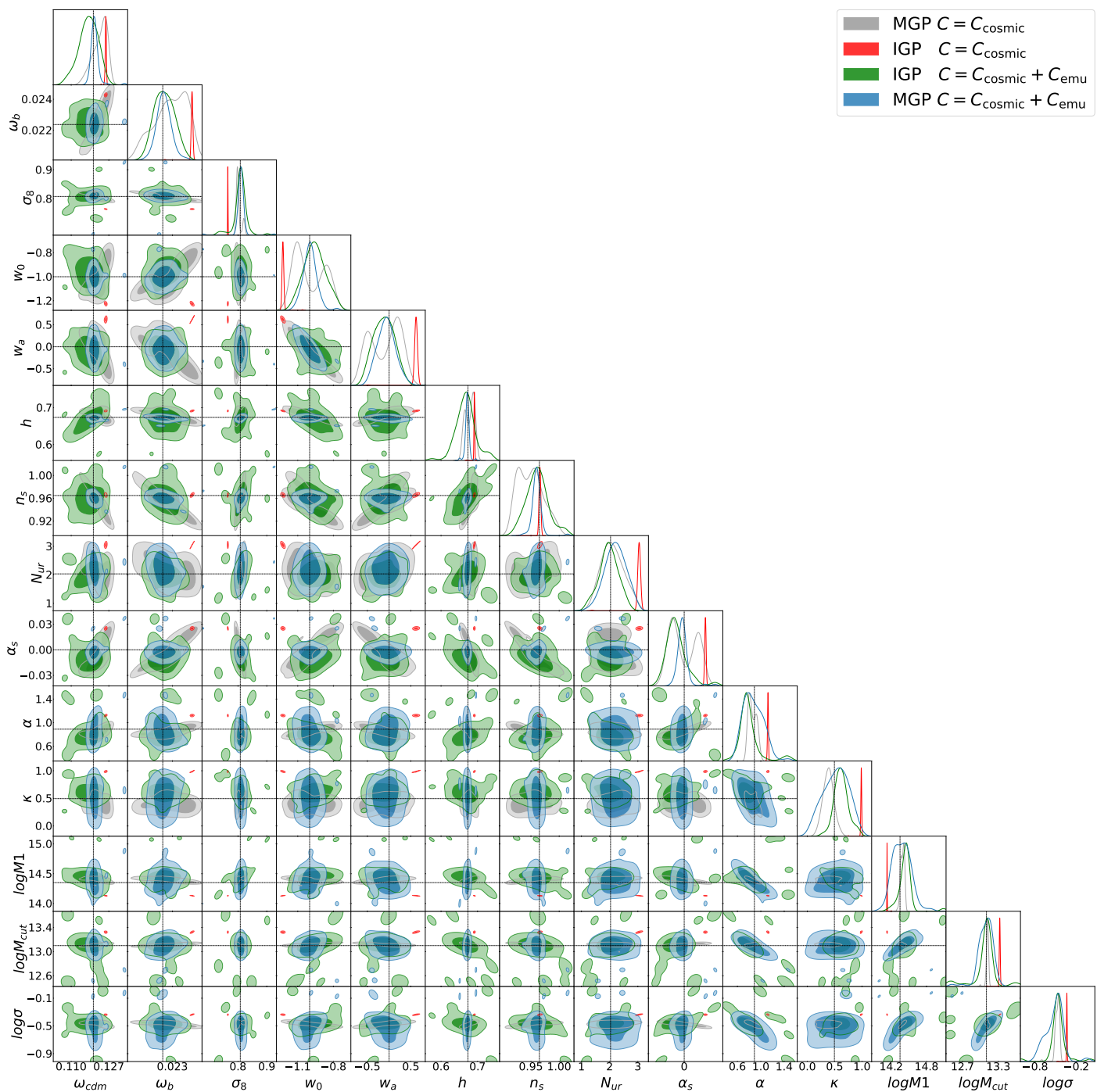}
	\caption{Posterior distribution of the cosmological and HOD parameters obtained after running MCMC inference on the average stack of the 25 correlation functions with Planck 2018 cosmology and the median HOD. The results obtained using the IGP model with(without) the model covariance prediction are shown in green(red), and those obtained using the MGP model with(without) the model covariance prediction are shown in blue(black). The true parameters are pointed by the black dotted lines. }
	\label{fig:mean_fit}
\end{figure*}



We first tested the performance of both IGP and MGP models to recover the parameters of interest when 
fitting a high precision measurement: the average stack of the 25 correlation function monopoles, corresponding to a volume of $200~(h^{-1} \mathrm{Gpc})^3$.
We used the likelihood defined in Eq.~\ref{eq:likelihood}, divided the sample covariance $C_\textrm{cosmic}$ by 25 and run two inferences with and without using the emulator predicted covariance $C_\textrm{emu}$, for each model.

Figure \ref{fig:mean_fit} shows the 68 and 95\% confidence regions obtained for the cosmological and HOD parameters. 
The true values are shown by the dotted lines.
We observe several interesting points. First, neglecting the model covariance  
leads to strongly biased results for the IGP with very sharp posterior distributions, 
while the MGP performs significantly better though yielding multimodal posteriors.
Second, using the model covariance allows to recover unbiased results for all cosmological parameters within one sigma for both models. 
Third, while MGP and IGP contours are consistent, the MGP model gives tighter constraints on cosmological parameters, and the IGP model gives tighter constraints on the HOD parameters. 
We also see in the posterior distributions several local maxima, specially in the green IGP contours. 
When sampling cosmological parameters far from the ones used in the training process of the GPs lead to larger model covariances while fitting the data points. 
This can also be seen as the goodness-of-fit term being counter-balanced by the determinant in the likelihood. 
Ignoring the model covariance can cause the fitter to converge toward a point that, when considering the total covariance, turns out to be only a local maxima in the log likelihood.


\begin{figure}
	\centering
	\includegraphics[width=1.\columnwidth]{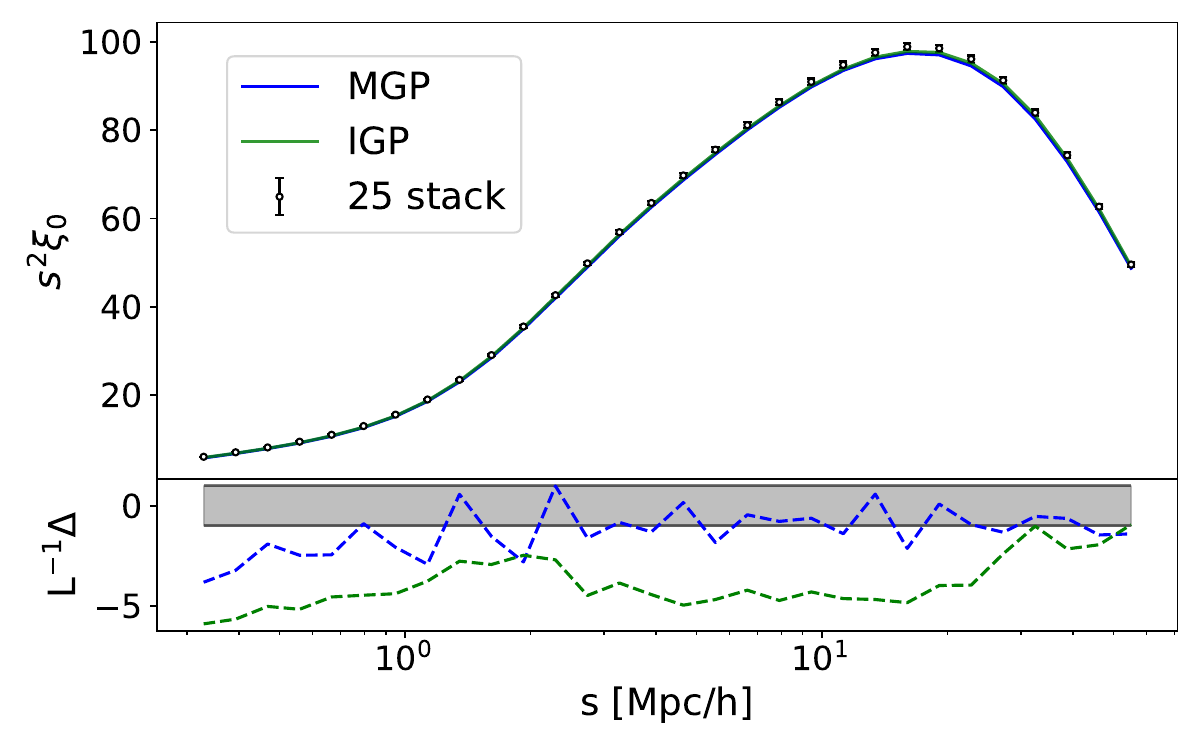}
	\caption{The 25 realisations of Planck2018 cosmology with the median HOD in grey. The average is shown in red dotted line and the MGP prediction in blue. The 25 residuals for both IGP (green) and MGP (blue) predictions are shown in the bottom panel. $L^{-1}$ is the Cholesky decomposition of the inverse of $C_\textrm{tot}$. }
	\label{fig:average_bestfit}
\end{figure}

Figure \ref{fig:average_bestfit} compares the best fits of the MGP and IGP models in blue and green respectively. The bottom panel displays the residuals, using the Cholesky decomposition matrix to account for correlations between scales. 
The one sigma deviation limit is set by the grey shaded area. 
For the MGP model we get a reduced chi squared of $\chi_r^2 = 5.6$ and for the IGP $\chi_r^2 = 31$. These large $\chi_r^2$ values are due to the large volume probed by the average of 25 simulations. These $\chi_r^2$ values and the bottom panel of Figure~\ref{fig:average_bestfit} show that the data are much better adjusted by the MGP model over all scales. 


\subsection{Cubic boxes : 25 individual fits}
\label{sec:reco:25}

\begin{table}
    \centering
    \caption{Summary statistics for the fits of the 25 cubic mocks with Planck2018 cosmology and the median HOD. $\langle \Delta_\textrm{p} \rangle$ is the bias, the difference between the true and best fitted value for any parameter $p$. $\langle \sigma_\textrm{p}\rangle$ is the average over the 25 realisations of the (symmetrised) one sigma confidence level.  $\textrm{std}\left( \textrm{p} \right )$ is the standard deviation of the estimated parameter $p$ over the 25 results. Note that $f$ is a derived parameter from the other cosmological ones.}
    {
    \begin{tabular}{l|ccc|ccc}
    \hline
    \hline
       & \multicolumn{3}{c|}{IGP $[\times 10^{-2}]$}  & \multicolumn{3}{c}{MGP $[\times 10^{-2} ]$} \\
     $p$ &
    $\langle \Delta_\textrm{p} \rangle$ & 
    $\langle \sigma_\textrm{p}\rangle$ & 
    $\textrm{std}\left( \textrm{p} \right )$ & 
    $\langle \Delta_\textrm{p} \rangle$ & 
    $\langle \sigma_\textrm{p}\rangle$ & 
    $\textrm{std}\left( \textrm{p} \right )$ \\

    \hline
    \hline
  $f$ & 0.06 & 1.88 & 0.33 & -0.14 & 1.80 & 0.20  \\ 
  \hline
  \hline
  $\omega_\textrm{m}$ & 0.05 & 0.50 & 0.14 & -0.02 & 0.16 & 0.04 \\ 
  $\omega_\textrm{b}$ & -0.02 & 0.09 & 0.02 & -0.01 & 0.06 & 0.02 \\ 
  $\sigma_8 $ & -0.28 & 1.64 & 0.34 & -0.05 & 0.36 & 0.04 \\ 
  $w_0$ & -1.61 & 10.96 & 2.79 & 0.21 & 4.85 & 0.43  \\ 
  $w_\textrm{a} $& 6.02 & 31.57 & 8.11& -0.22 & 19.04 & 2.57  \\ 
  $h $& 0.15 & 2.33 & 0.45& 0.08 & 0.65 & 0.12  \\ 
  $n_\textrm{s}$ & 0.43 & 2.53 & 0.63 & 0.11 & 0.72 & 0.16 \\ 
  $N_\textrm{ur}$& -1.15 & 41.97 & 8.31 & -5.65 & 46.56 & 8.02  \\ 
  $\alpha_\textrm{s}$& 0.56 & 1.36 & 0.71 & 0.02 & 0.48 & 0.08  \\

    \hline
    \hline 

  $\alpha$ & 11.10 & 11.24 & 6.35& -0.38 & 19.03 & 4.19  \\ 
  $\kappa$ & -10.20 & 20.31 & 6.42 & -6.70 & 27.70 & 4.27 \\ 
  $\log M_1$ & -7.67 & 8.86 & 4.69& 0.39 & 17.39 & 3.09  \\ 
  $\log M_\textrm{cut}$ & 0.45 & 4.99 & 2.24& 4.86 & 10.56 & 2.83  \\ 
  $\log \sigma$ & -1.09 & 5.70 & 4.08 & 3.78 & 13.50 & 3.86  \\ 
 
    \hline
    \hline
    
    \end{tabular}
    }
    \label{tab:stats_cubic}
\end{table}

\begin{figure}
  \centering
 \includegraphics[width=0.7\columnwidth]{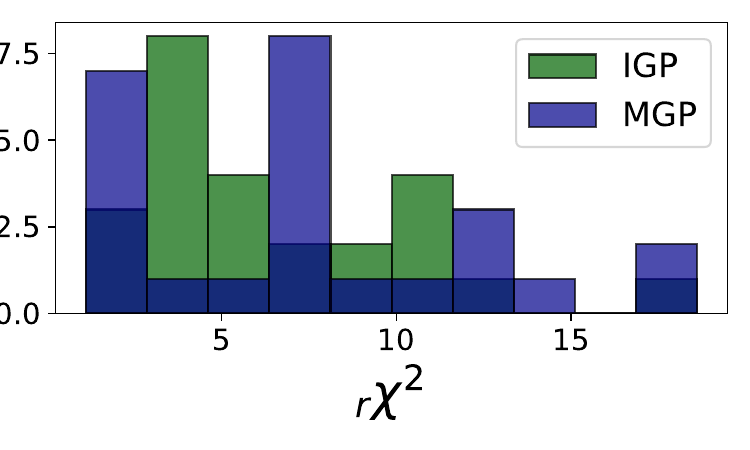}  
  \includegraphics[width=0.7\columnwidth]{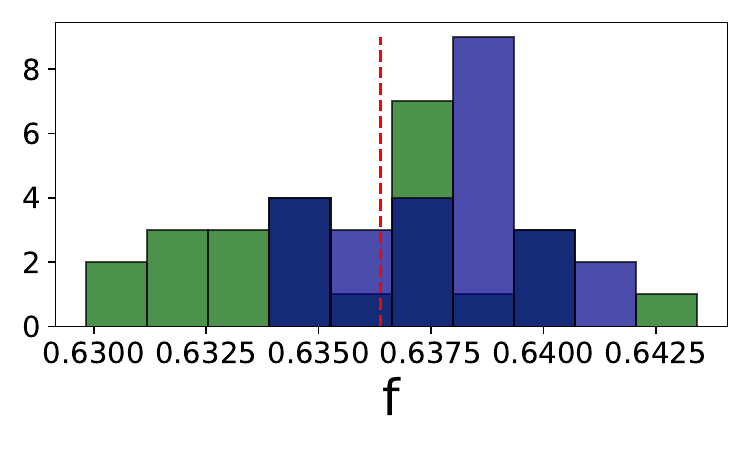}  
  \includegraphics[width=0.7\columnwidth]{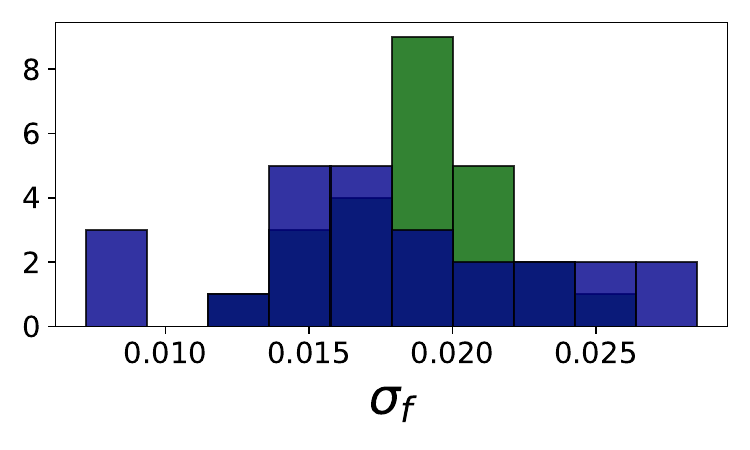}
\caption{ Reduced chi squared, best fitted growth rate and corresponding uncertainties, resulting from the separate fits of the 25 cubic mocks with Planck2018 cosmology and the median HOD. Results using the MGP and IGP models are shown in blue and green respectively. The true parameters values are pointed by a red dotted line.}
\label{fig:25_box_GrowthRate}
\end{figure}

\begin{figure*}
  \centering
  \includegraphics[width=0.195\textwidth]{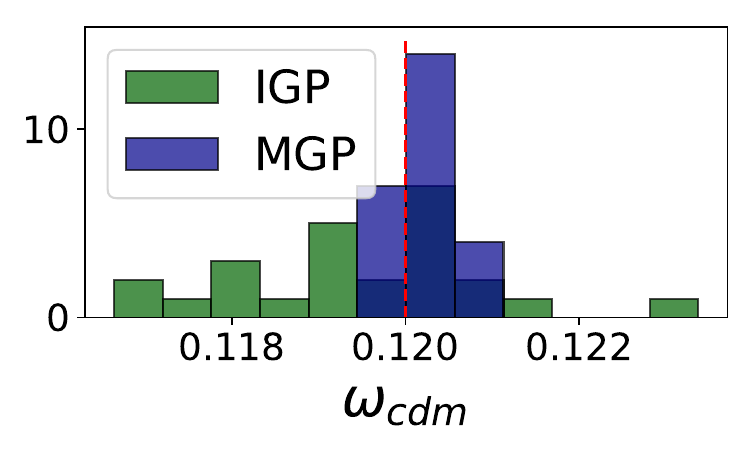}  
  \includegraphics[width=0.195\textwidth]{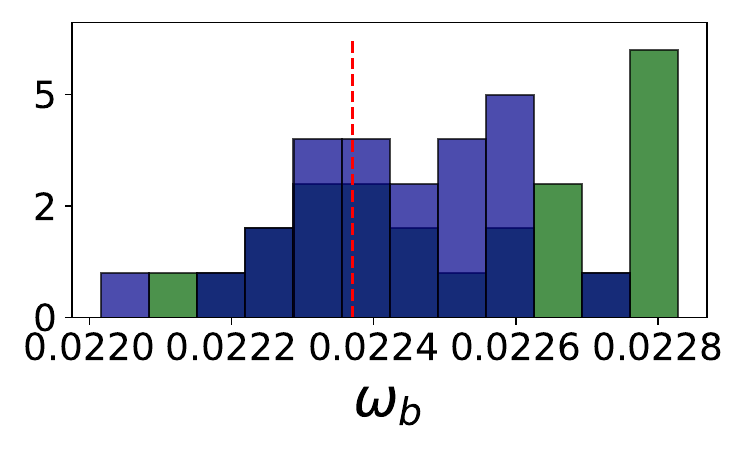}  
  \includegraphics[width=0.195\textwidth]{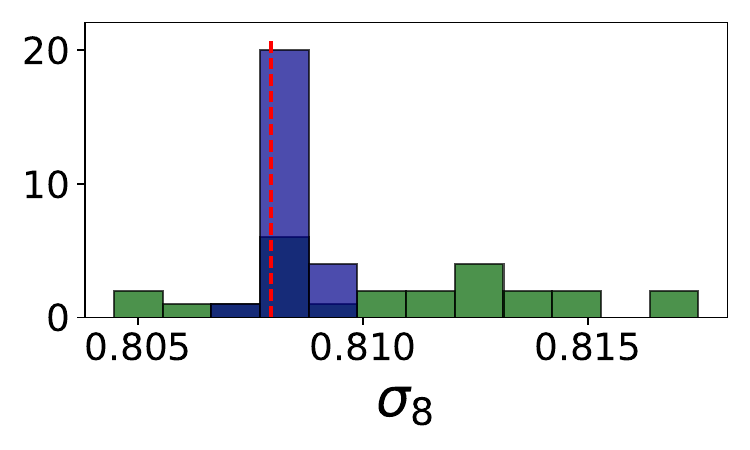} 
  \includegraphics[width=0.195\textwidth]{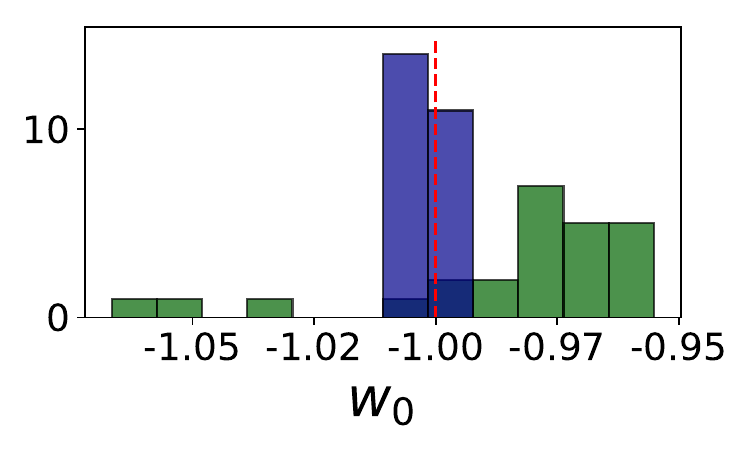}  
  \includegraphics[width=0.195\textwidth]{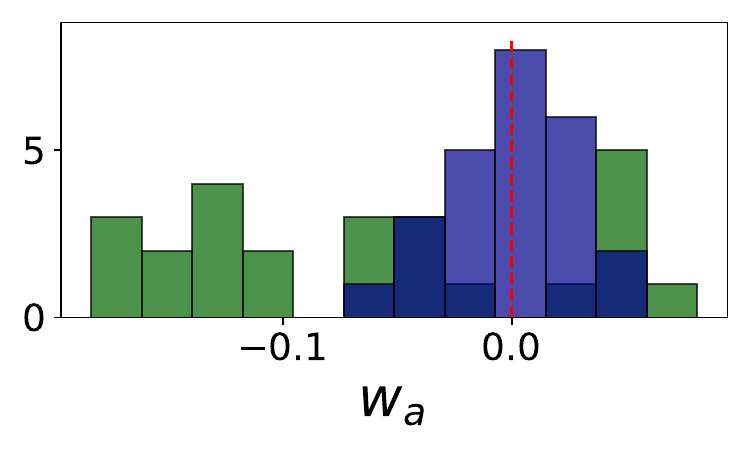}  
  \includegraphics[width=0.195\textwidth]{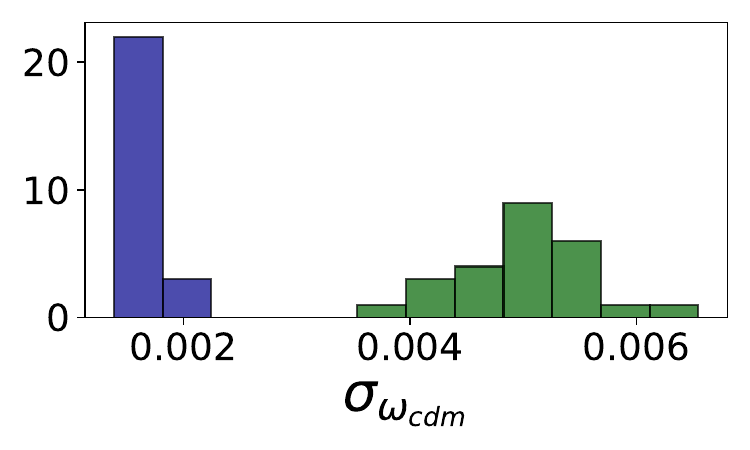}  
  \includegraphics[width=0.195\textwidth]{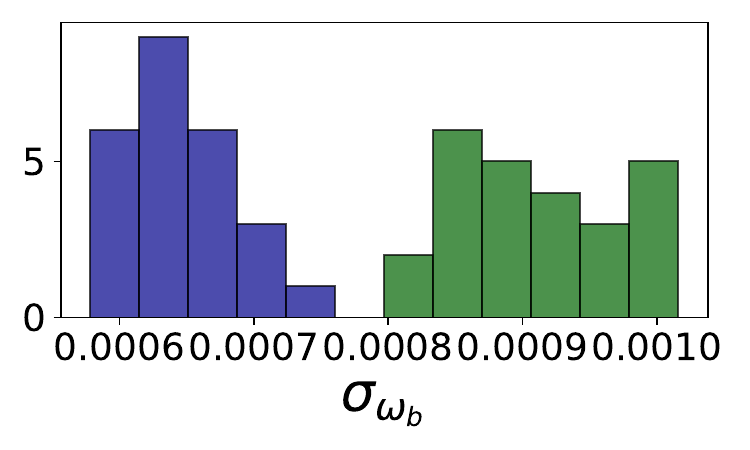}  
  \includegraphics[width=0.195\textwidth]{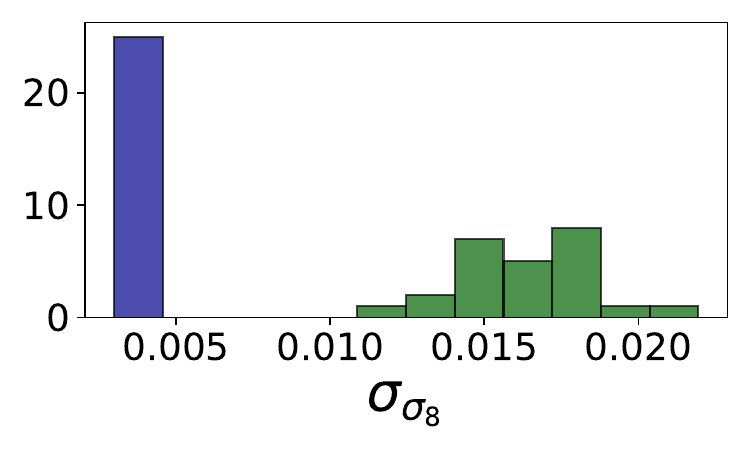}
  \includegraphics[width=0.195\textwidth]{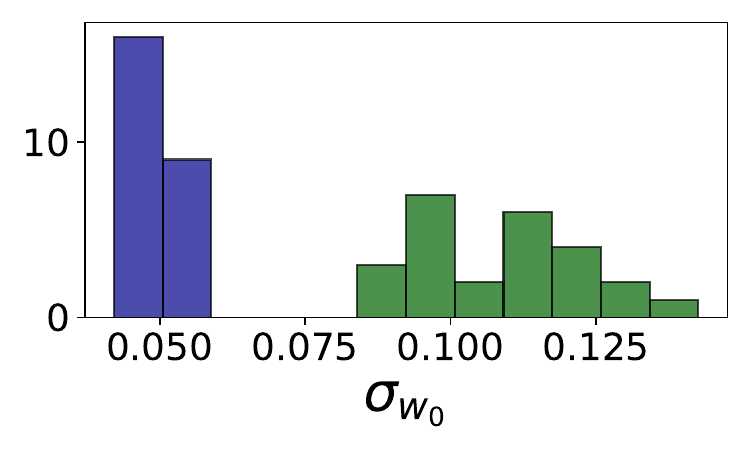}  
  \includegraphics[width=0.195\textwidth]{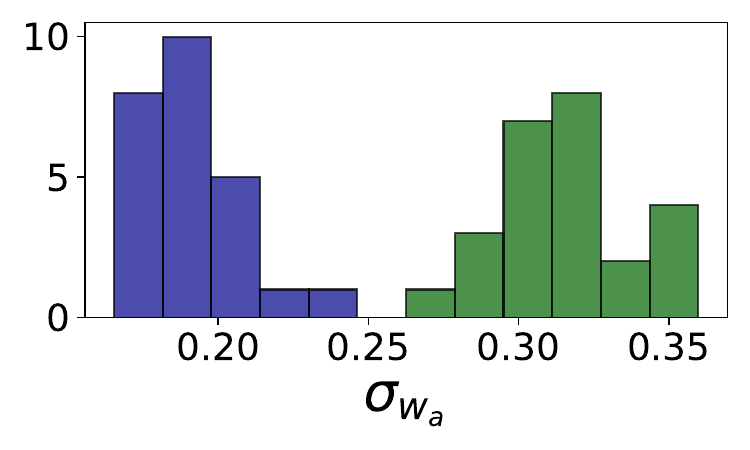}  
  \hfill 
  
  {\hfill
  \includegraphics[width=0.195\textwidth]{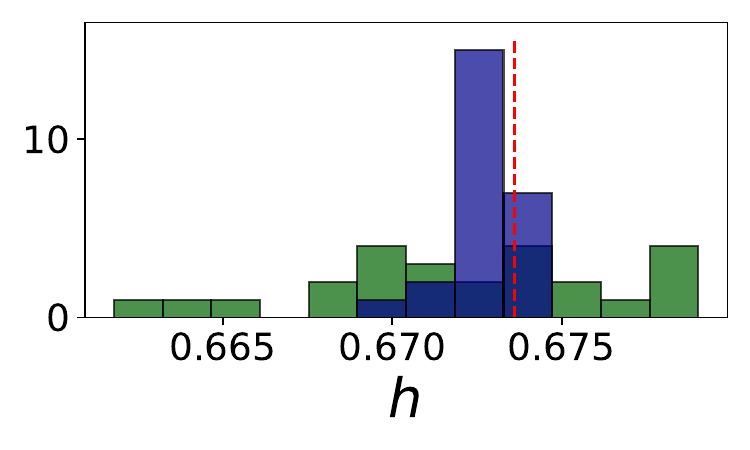}  
  \includegraphics[width=0.195\textwidth]{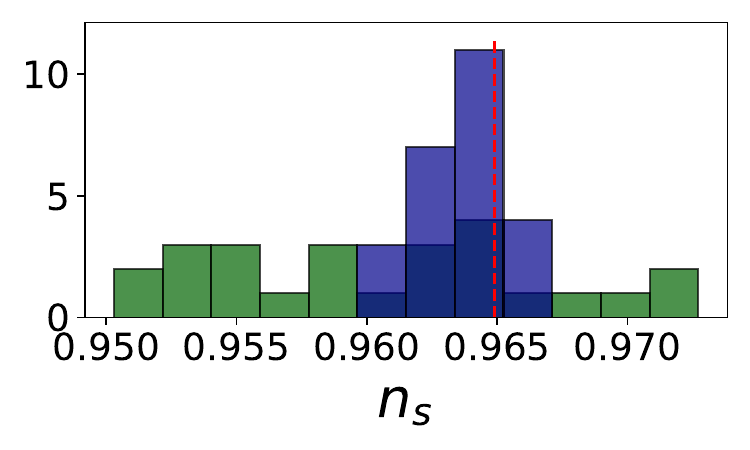}  
  \includegraphics[width=0.195\textwidth]{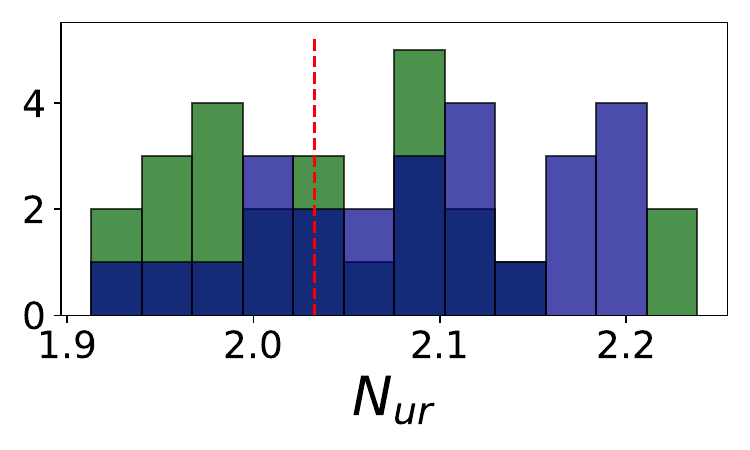}  
  \includegraphics[width=0.195\textwidth]{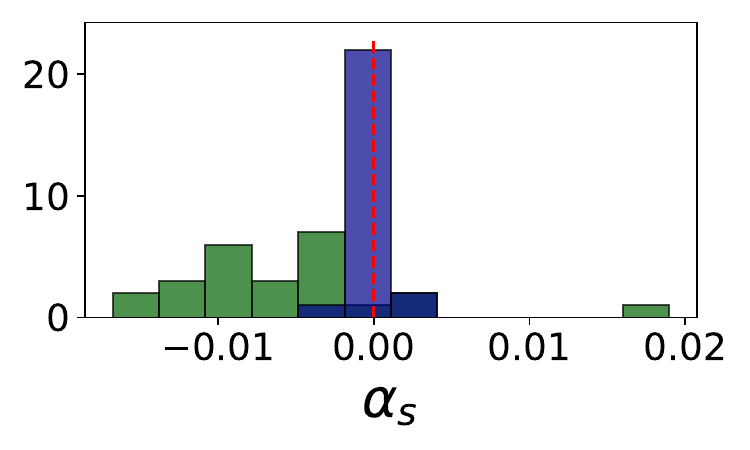}  
  \hfill}

 {\hfill
  \includegraphics[width=0.195\textwidth]{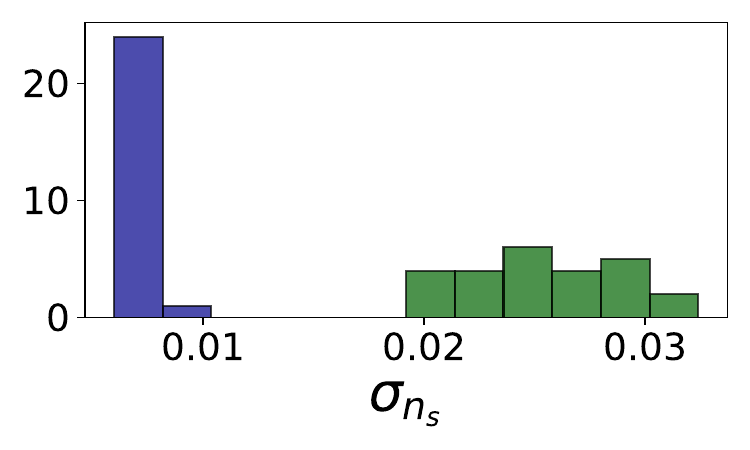}  
  \includegraphics[width=0.195\textwidth]{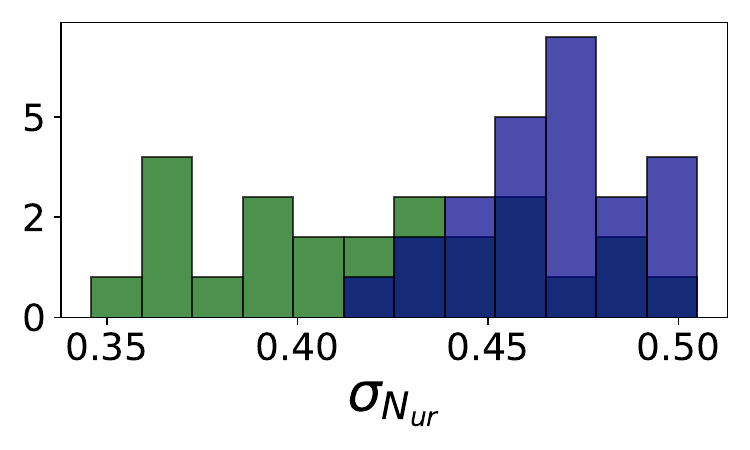}  
  \includegraphics[width=0.195\textwidth]{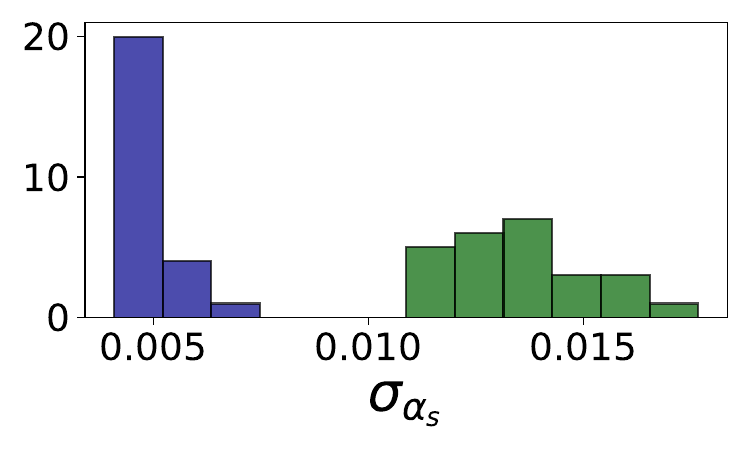}  
  \includegraphics[width=0.195\textwidth]{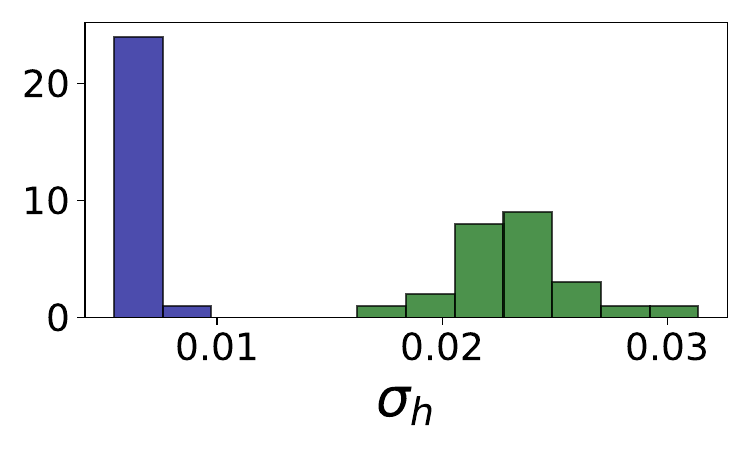}  
  \hfill}
  
\caption{Best fitted cosmological parameters along with the corresponding uncertainties, on the 25 cubic mocks with Planck2018 cosmology and the median HOD. Results using the MGP and IGP models are shown in blue and green respectively.The true parameters values are pointed by a red dotted line. }
\label{fig:25_box_cosmo}
\end{figure*}

\begin{figure*}
  \centering
  \includegraphics[width=0.195\textwidth]{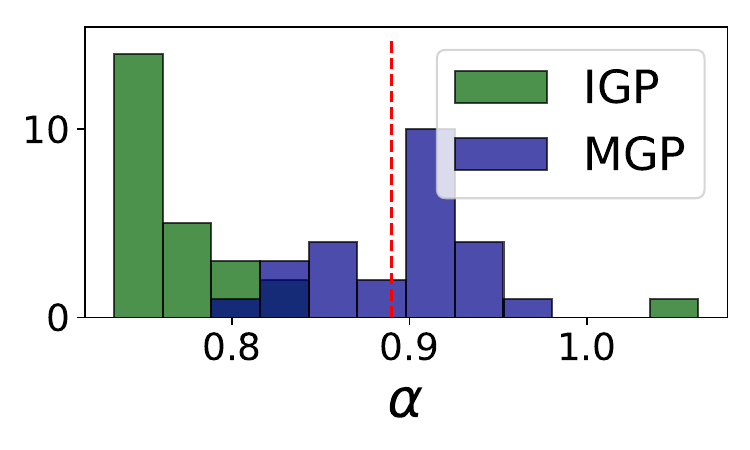}  
  \includegraphics[width=0.195\textwidth]{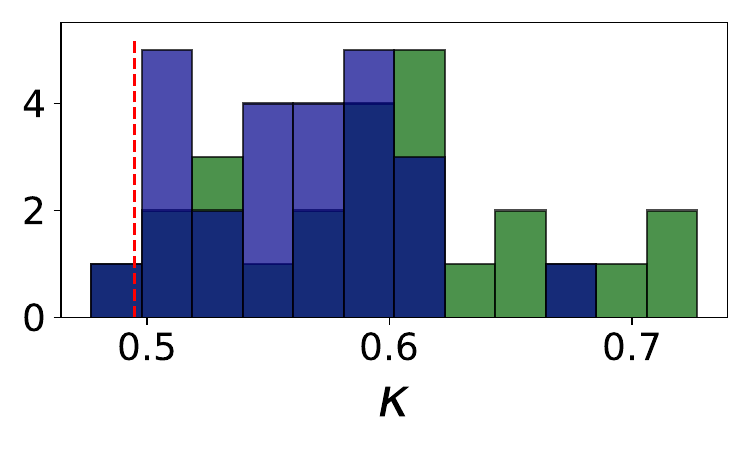}  
  \includegraphics[width=0.195\textwidth]{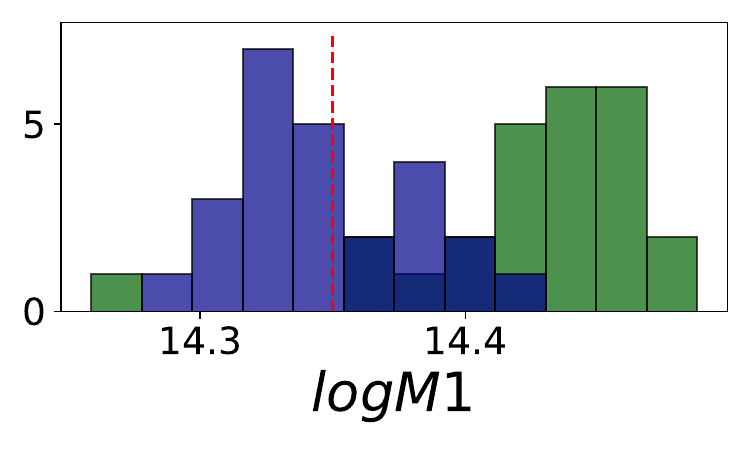} 
    \includegraphics[width=0.195\textwidth]{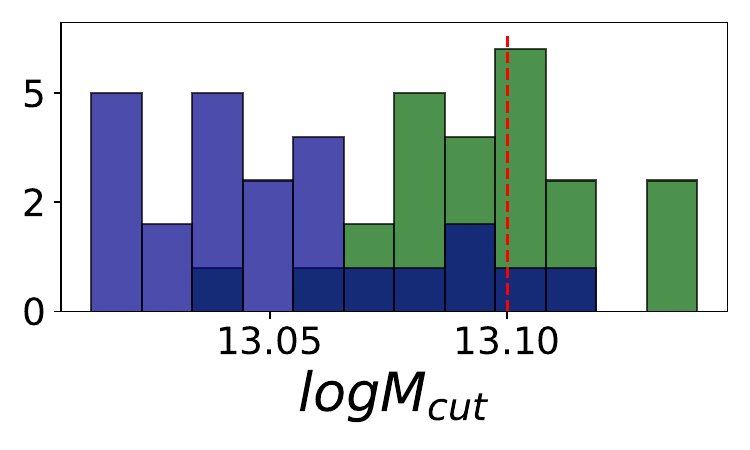}  
  \includegraphics[width=0.195\textwidth]{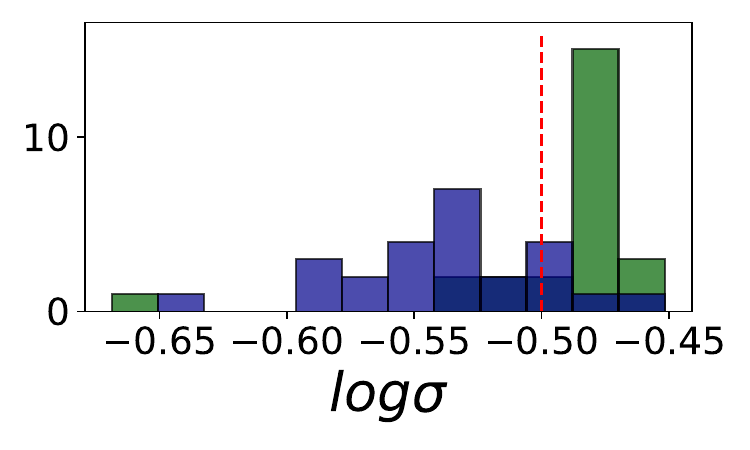} 
  \includegraphics[width=0.195\textwidth]{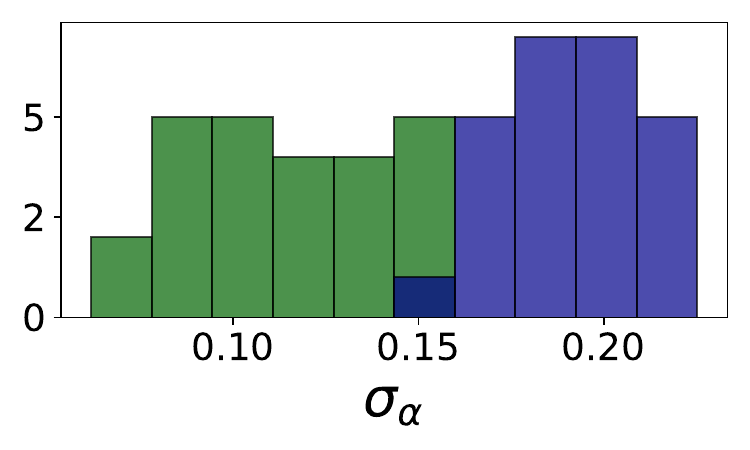}  
  \includegraphics[width=0.195\textwidth]{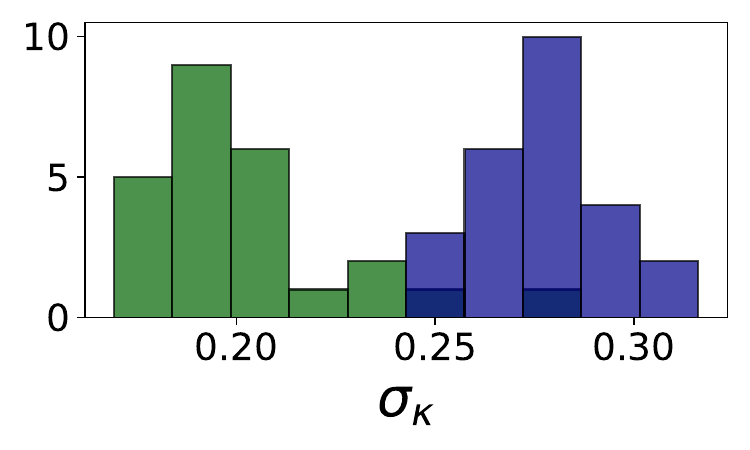}  
  \includegraphics[width=0.195\textwidth]{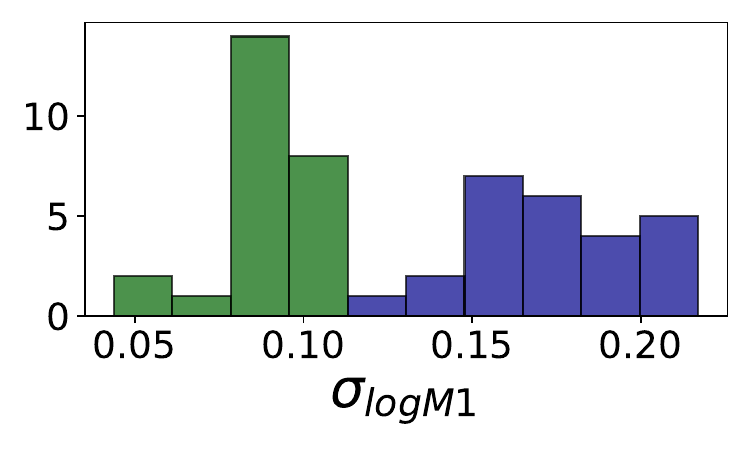} 
    \includegraphics[width=0.195\textwidth]{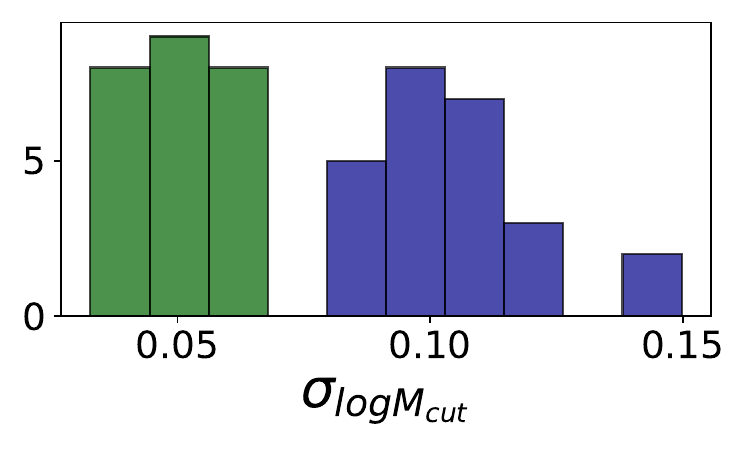}  
  \includegraphics[width=0.195\textwidth]{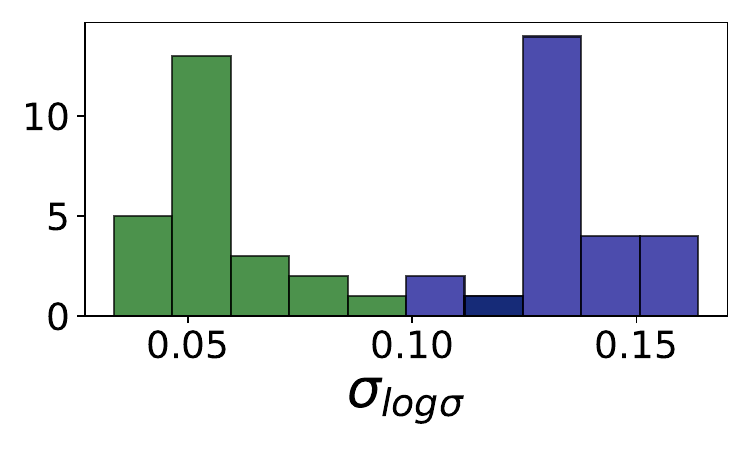}  

\caption{Best fitted HOD parameters along with the corresponding uncertainties, on the 25 cubic mocks with Planck2018 cosmology and the median HOD. Results using the MGP and IGP models are shown in blue and green respectively. The true parameters values are pointed by a red dotted line.}
\label{fig:25_box_HOD}
\end{figure*}

To  check statistically the accuracy and potential biases 
of our models, we fit each of the 25 boxes separately, corresponding to a probed volume comparable to current surveys.
The cosmic covariance matrices were normalised to account for this change in volume.
Figure \ref{fig:25_box_GrowthRate} shows the reduced chi-squared $\chi_r^2$, the best-fit growth rate parameter $f$ along with their estimated uncertainties $\sigma_f$. 
The derived cosmological parameters along with their inferred uncertainties are shown in Figure~\ref{fig:25_box_cosmo} and the same for the HOD parameters in Figure~\ref{fig:25_box_HOD}. 
Table \ref{tab:stats_cubic} reports some summary statistics from those results, where $\langle \Delta_p \rangle$ is the difference between the true and best-fit 
value for parameter $p$; 
$\langle \sigma_p\rangle$  is the average over the 25 realisations of the (symmetrised) one sigma confidence level; and $\textrm{std}(p)$ is the standard deviation of the estimated parameter $p$ over the 25 fits. 
We note that the bias in the growth rate measurement is reduced with our MGP model, and that both models give consistent results, recovering the growth rate within one sigma. 
The reduced chi-squared distributions are similar, but with this sample the IGP model gives in average a better fit with $\langle \chi_r^2 \rangle =  4.93 $ while we have $\langle \chi_r^2 \rangle = 7.36 $ for MGP for $\mathrm{dof} = (30 - 14) = 16$ degrees of freedom.
Although it is hard to draw conclusions because of the limited size of the sample, 
in both cases the mean estimated error $\langle \sigma_f \rangle$ does not exactly 
match the standard deviation of the best-fit parameters $\textrm{std}( p )$, 
but is conservatively larger. 
We also see that our MGP model yields smaller dispersion and in tighter constraints on $f$.

Figure \ref{fig:25_box_cosmo} showcases similar results for the cosmological parameters. 
Both models are able to retrieve the right parameters within one sigma, 
but the MGP returns smaller bias, smaller dispersion and tighter constraints 
for all parameters except the number of ultra relativistic species $N_{\rm ur}$ which is poorly constrained. 
Specially, the uncertainty on amplitude parameter $\sigma_8$ is reduced by a factor larger than 4, while reducing the bias on the best fitted value.
For the HOD parameters illustrated in Figure~\ref{fig:25_box_HOD}, 
we likewise get unbiased measurements. 
Once again the bias and the dispersion for the best fitted values are overall reduced with the MGP, however the IGP gets tighter constraints. Hence, it seems that using the correlation between scales helps constraining the cosmology rather than the HOD parameters as the HOD signal is mainly in the small scales where the cosmic covariance is nearly diagonal (see Figure~\ref{fig:cov}), while all scales are affected by cosmological parameters. 

%


  

\subsection{Observational effects : wide angle and AP}
\label{sec:reco:AP}

We validated the MGP model on cubic mocks similar to the simulations used to train the 
emulators and showed that it performs significantly better than the standard IGP. Now we wish to test the constraining power and accuracy of the MGP model
on more realistic measurements.

For each of the 25 realisations, we placed the observer at the center of the box. 
As the snapshot describes the clustering at redshift $z=0.2$ we select galaxies at 
redshifts $0.15 < z < 0.25$ from the observer, in order to not make large changes to the observed clustering. We consider each survey to cover the full sky. 
We used the Landy-Szalay estimator of Eq.~\ref{eq:LS} to compute correlation functions. 
The measurement of the 2PCF requires the choice of a specific fiducial cosmology to 
convert redshifts into comoving distances. 
Differences between the fiducial and the true cosmology lead to distortions 
on the measured clustering known as the Alcock-Paczynski (AP) effect \citep{alcock-paczynski}.
This effect is often neglected when building an emulator \citep{yuan_stringent_2022, zhai_aemulus_2022} 
as it is small on very small scales. 
Moreover, modeling the AP distortion requires either one extra parameter in the training input space or to have a continuous model over different scales for interpolation. 
By construction, our MGP model is more adapted than the 
IGP to build a continuous model, which can be evaluated at any scale within the training range.

If the choice of fiducial cosmology is not far from the true cosmology, the AP effect can be modelled with two parameters scaling the radial and angular separations.
However, since we are only using the monopole of the 2PCF, we only have access to isotropic information. We use the isotropic dilatation parameter $\alpha_V$ described in \cite{moon_first_2023} to scale separations as :
\begin{equation}
\xi_0^\textrm{obs}(s) = \xi_0^\textrm{mod}(\alpha_V s)
\label{eq:AP_Rescaling}
\end{equation}
where 
\begin{equation}
\alpha_V  = \frac{D_V(z)}{D^\textrm{fid}_V(z)}
\label{eq:AP_alpha}
\end{equation}
with the isotropic distance defined as a combination of the Hubble distance (radial) 
$D_H(z) = c/H(z)$ and the comoving angular distance $D_M(z) = (1+z)D_A(z)$ :
\begin{equation}
D_V(z) = \left [ z D_H(z)D_M^2(z) \right]^{1/3}.
\label{eq:AP_Dv}
\end{equation}

\begin{figure}
	\centering
	\includegraphics[width=1.\columnwidth]{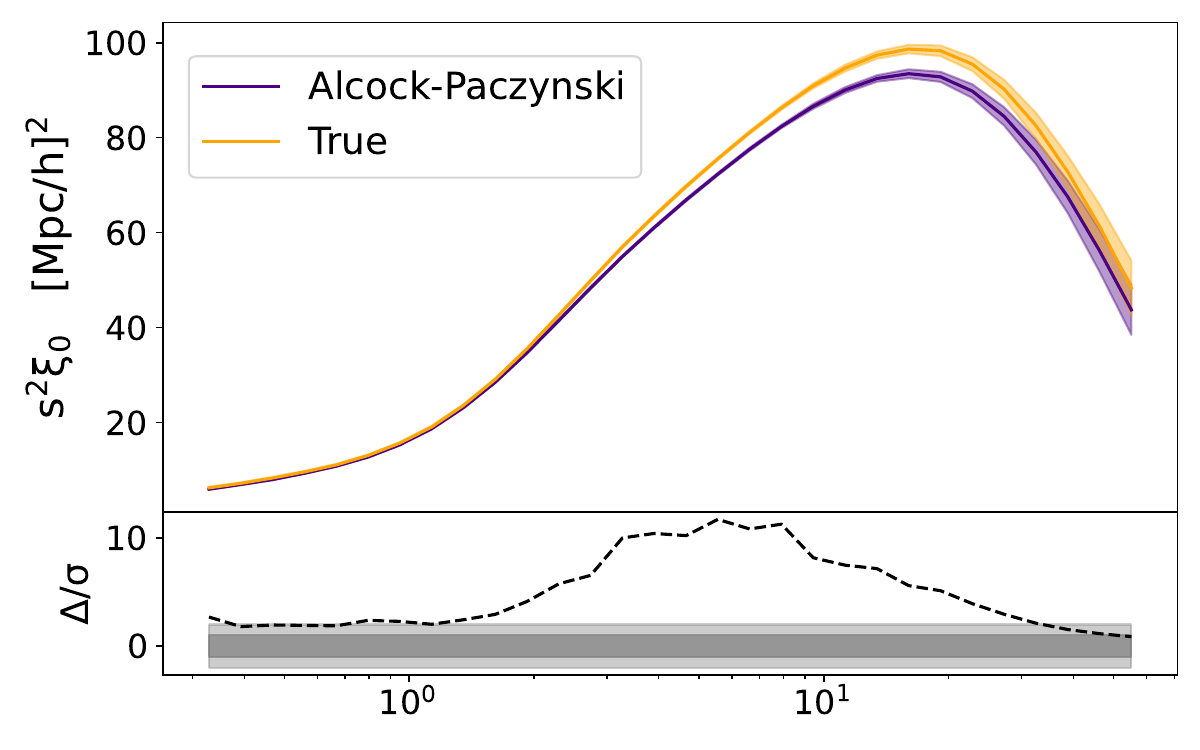}
	\caption{Top panel : mean clustering with its standard deviation of the 25 mocks with Planck2018 cosmology, the median HOD and a spherical full sky footprint from $z = 0.15$ to $z=0.25$. The true and AP distorted clustering are shown in orange and indigo respectively. The bottom panel shows the residual using the cosmic variance of the varying LoS clustering.}
	\label{fig:ap_effect}
\end{figure}

\begin{table}
    \centering
    \caption{Summary statistics for the fits of the 25 mocks with spherical full sky footprint from $z = 0.15$ to $z = 0.25$, Planck2018 cosmology and the median HOD, where the clustering was measured using cosmology \textsc{c177} of the \textsc{AbacusSummit} suite as fiducial cosmology resulting in  an isotropic dilatation parameter $\alpha_v \approx 1.02$. $\langle \Delta_\textrm{p} \rangle$ is the bias, the difference between the true and best fitted value for any parameter $p$. $\langle \sigma_\textrm{p}\rangle$ is the average over the 25 realisations of the (symmetrised) one sigma confidence level.  $\textrm{std}\left( \textrm{p} \right )$ is the standard deviation of the estimated parameter $p$ over the 25 results. Note that $f$ is a derived parameter from the other cosmological ones.}
    {
    \begin{tabular}{l|ccc|ccc}
    \hline
    \hline
      & \multicolumn{3}{c|}{MGP$_{\alpha_V}$  $[\times 10^{-2}]$}  & \multicolumn{3}{c}{MGP$[\times 10^{-2}]$} \\
       & \multicolumn{3}{c|}{accounting $\alpha_V$}  & \multicolumn{3}{c}{neglecting for $\alpha_V$} \\
     $p$ &
    $\langle \Delta_\textrm{p} \rangle$ & 
    $\langle \sigma_\textrm{p}\rangle$ & 
    $\textrm{std}\left( \textrm{p} \right )$ & 
    $\langle \Delta_\textrm{p} \rangle$ & 
    $\langle \sigma_\textrm{p}\rangle$ & 
    $\textrm{std}\left( \textrm{p} \right )$ \\

    \hline
    \hline

  $f$ &-0.02 & 0.59 & 0.10& -0.20 & 0.75 & 0.15  \\ 
  \hline
  \hline
  $\omega_\textrm{m}$ &-0.00 & 0.10 & 0.02 & -0.04 & 0.13 & 0.03 \\ 
  $\omega_\textrm{b}$ &-0.00 & 0.04 & 0.00& -0.00 & 0.05 & 0.01  \\ 
  $\sigma_8 $ &-0.04 & 0.19 & 0.01& -0.04 & 0.23 & 0.02  \\ 
  $w_0$  &-0.13 & 2.93 & 0.28& -0.03 & 3.27 & 0.41  \\ 
  $w_\textrm{a} $  &0.56 & 11.77 & 0.80& 0.34 & 12.68 & 1.67 \\ 
  $h $ &0.02 & 0.42 & 0.08 & 0.15 & 0.52 & 0.09 \\ 
  $n_\textrm{s}$ &0.01 & 0.45 & 0.07& 0.01 & 0.50 & 0.08  \\ 
  $N_\textrm{ur}$  &-1.34 & 30.49 & 2.05& -2.94 & 31.75 & 5.01 \\ 
  $\alpha_\textrm{s}$ &0.00 & 0.22 & 0.02 & 0.01 & 0.28 & 0.04 \\ 

  \hline
  \hline
  
  $\alpha$  &5.66 & 32.97 & 5.97& 10.54 & 23.24 & 4.75 \\ 
  $\kappa$ &-1.57 & 26.37 & 1.23& 0.04 & 22.84 & 4.53  \\ 
  $\textrm{log}M_1$ &-0.51 & 32.20 & 3.99& -1.27 & 30.42 & 8.47  \\ 
  $\textrm{log}M_\textrm{cut}$  &5.06 & 11.15 & 2.09& 8.95 & 11.12 & 2.87 \\ 
  $\textrm{log} \sigma$ &6.96 & 15.94 & 2.92 & 4.74 & 13.97 & 2.56 \\

    \hline
    
    \end{tabular}
    }
    \label{tab:stats_ap}
\end{table}

We picked cosmology \texttt{c177} of the \textsc{AbacusSummit} suite as a fiducial cosmology which is different than the true one (Planck18),
resulting in a $2 \%$ variation in the isotropic dilatation parameter 
$\alpha_v \approx 1.02$. 
%
%
%
We measure the clustering of the 25 realisations of our survey-like mocks
using both the true and c177 as fiducial cosmologies.
The average monopoles along with the residual (using the cosmic variance of the survey-like mocks) are shown in Figure \ref{fig:ap_effect}.
Although the cosmic variance is larger for the survey-like mocks compared to the 
cubic boxes measurements due to the reduced volume, the AP distortions lead to 
significant deviations for every scales, up to 10 $\sigma$ for scales of few $\hmpc$.

To assess if ignoring the AP distortions can lead to biased cosmological constraints, we fit separately the 25 realisations of the AP distorted clustering, using the MGP model with and without (setting $\alpha_V = 1$) accounting for the 
AP distortions.
We refer to the former as MGP$_{\alpha_V}$.
In both cases we rescale the sample covariance matrix with the corresponding spherical volume. 
We also compare the impact of ignoring the AP effect using MGP and IGP in appendix \ref{sec:annex:ap}.

\begin{figure}
  \centering
 \includegraphics[width=0.8\columnwidth]{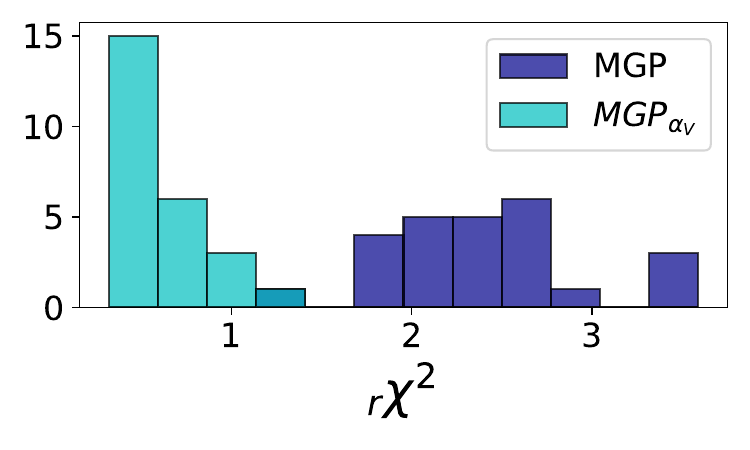}  
  \includegraphics[width=0.8\columnwidth]{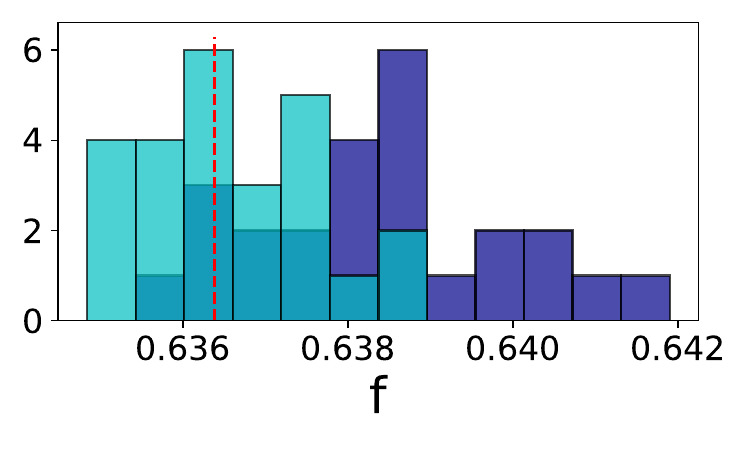}  
  \includegraphics[width=0.8\columnwidth]{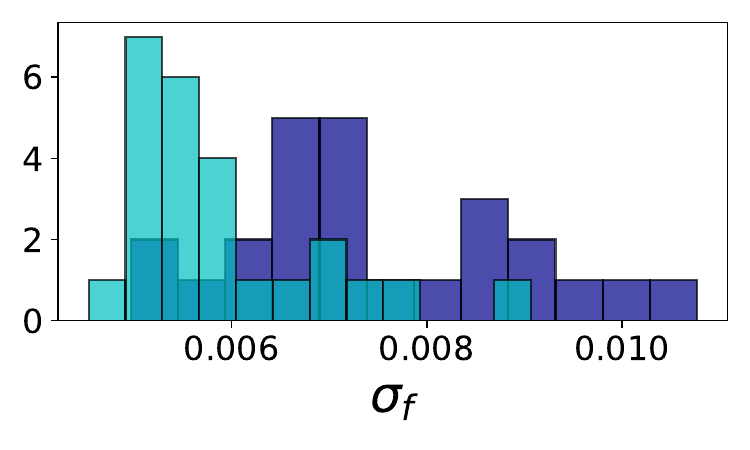}
\caption{Reduced chi squared, best fitted growth rate and corresponding uncertainties, resulting from the separate fits of the 25 mocks with spherical full sky footprint from $z = 0.15$ to $z = 0.25$, AP distortion, Planck2018 cosmology and the median HOD. Results using the MGP with and without modeling AP distortions are shown in cyan and blue respectively. The true growth rate value is pointed by a red dotted line.}
\label{fig:25_ap_chi2}
\end{figure}

\begin{figure*}
  \centering
  {\hfill
  \includegraphics[width=0.195\textwidth]{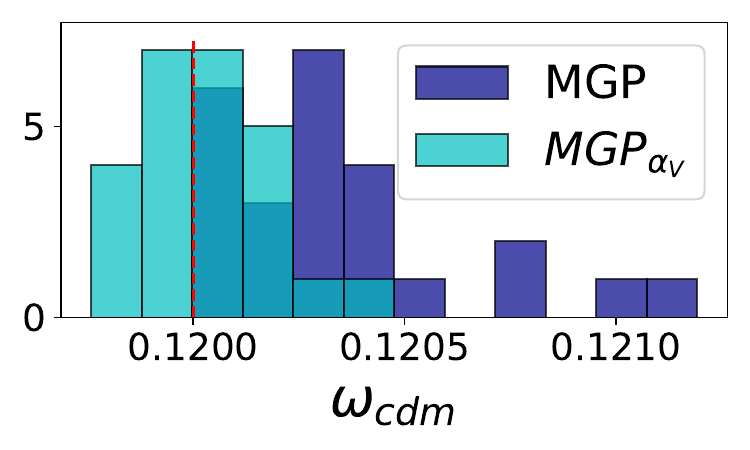}  
  \includegraphics[width=0.195\textwidth]{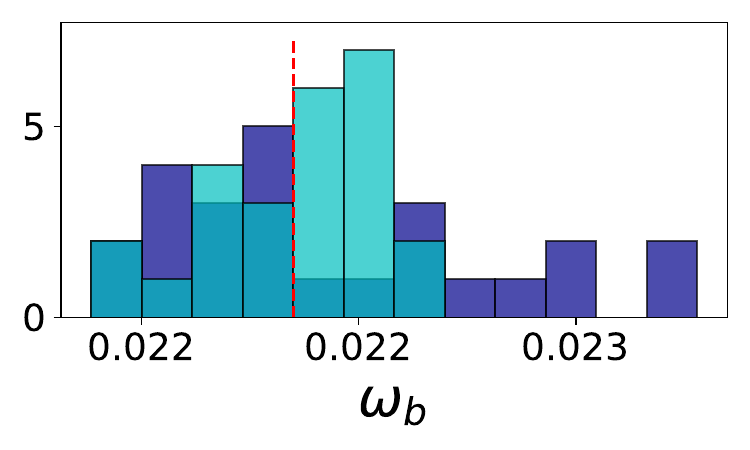}  
  \includegraphics[width=0.195\textwidth]{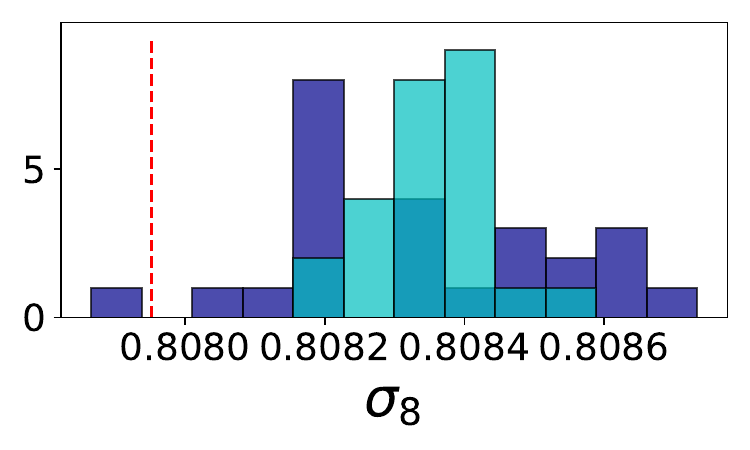} 
    \includegraphics[width=0.195\textwidth]{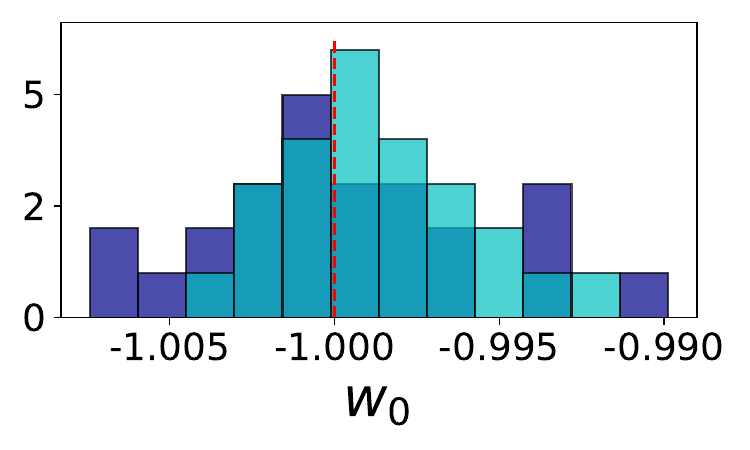}  
  \includegraphics[width=0.195\textwidth]{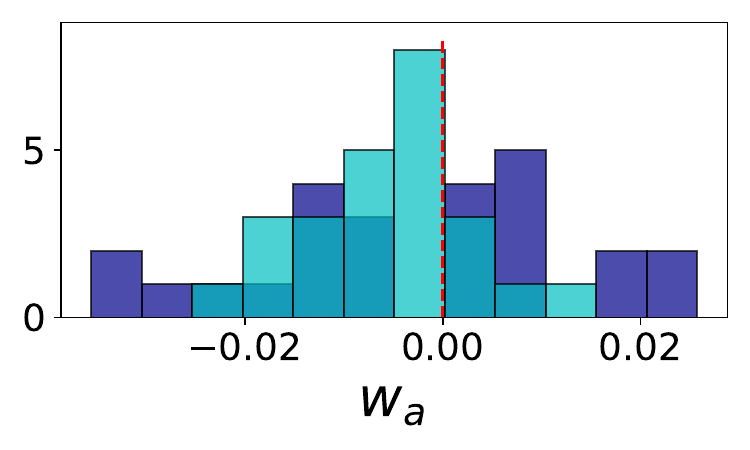} 
  \includegraphics[width=0.195\textwidth]{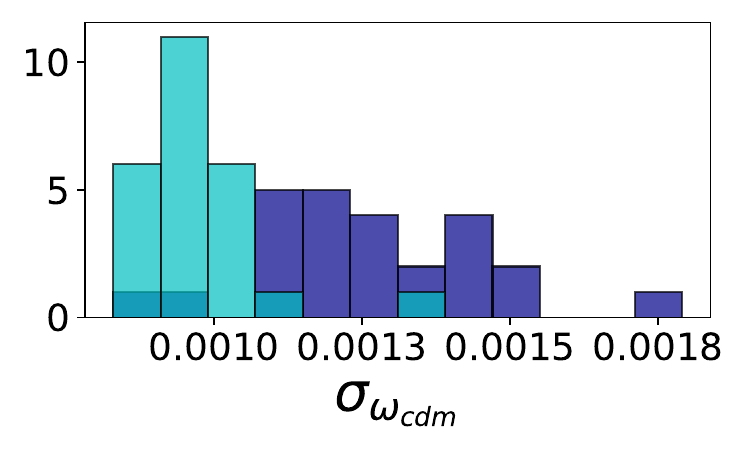}  
  \includegraphics[width=0.195\textwidth]{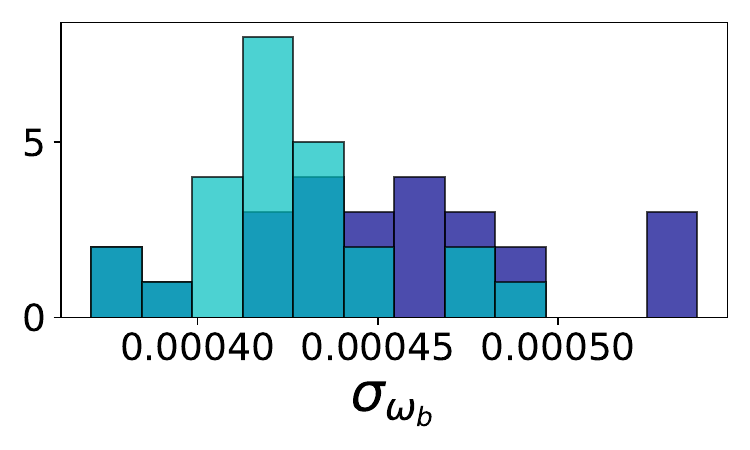}  
  \includegraphics[width=0.195\textwidth]{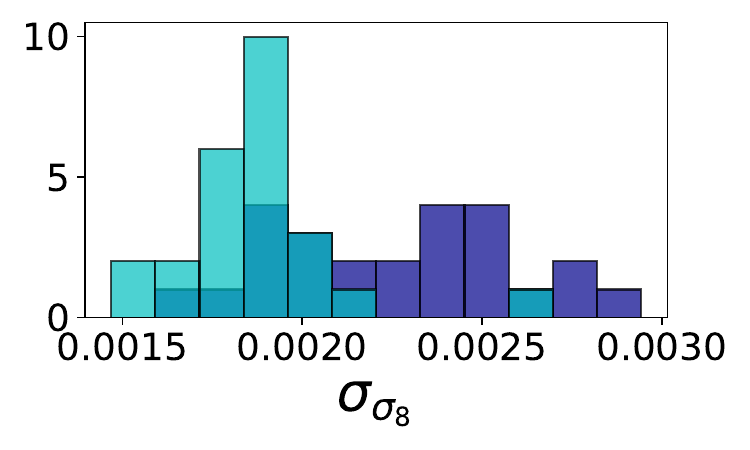}
    \includegraphics[width=0.195\textwidth]{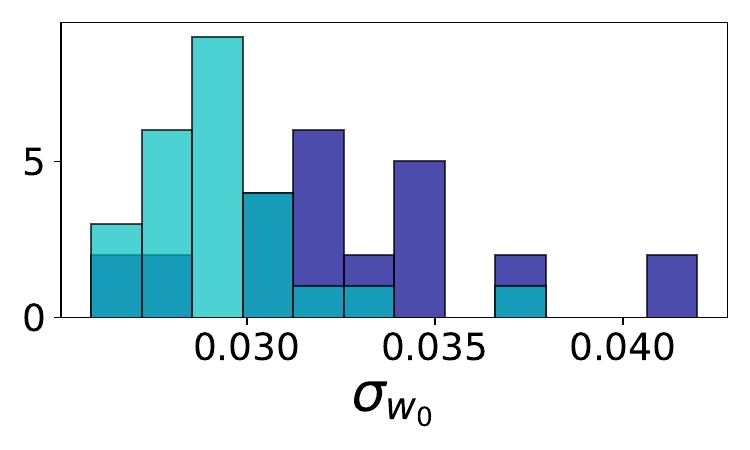}
  \includegraphics[width=0.195\textwidth]{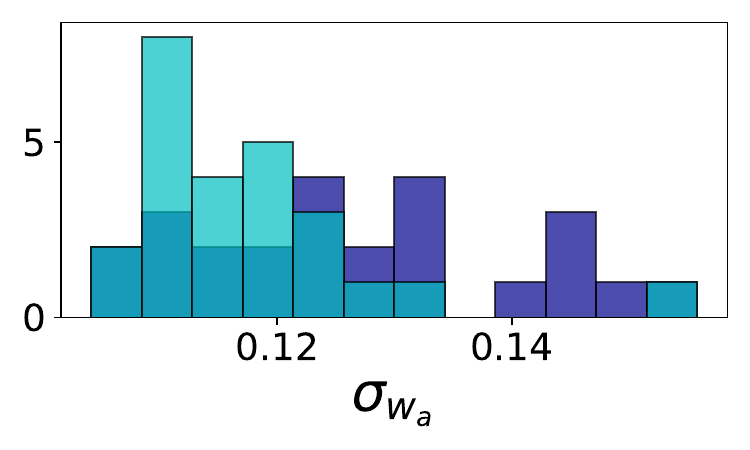} 
  \hfill}
  
  {\hfill
    \includegraphics[width=0.195\textwidth]{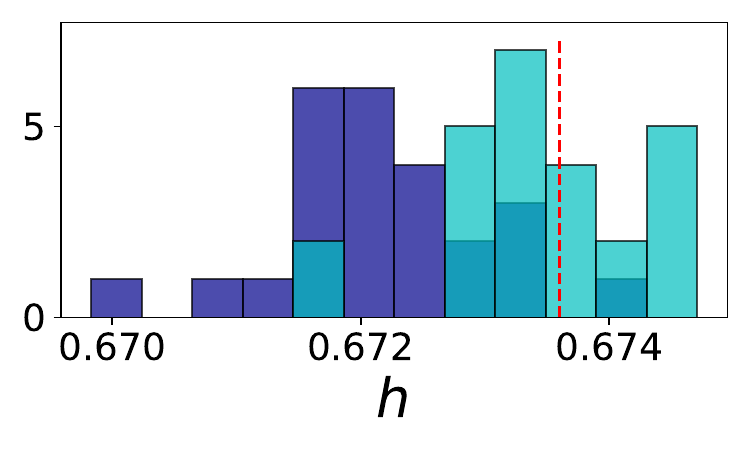}  
    \includegraphics[width=0.195\textwidth]{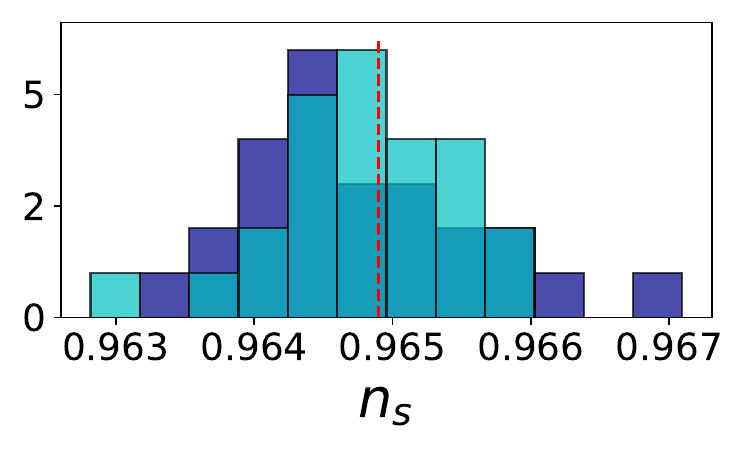}  
  \includegraphics[width=0.195\textwidth]{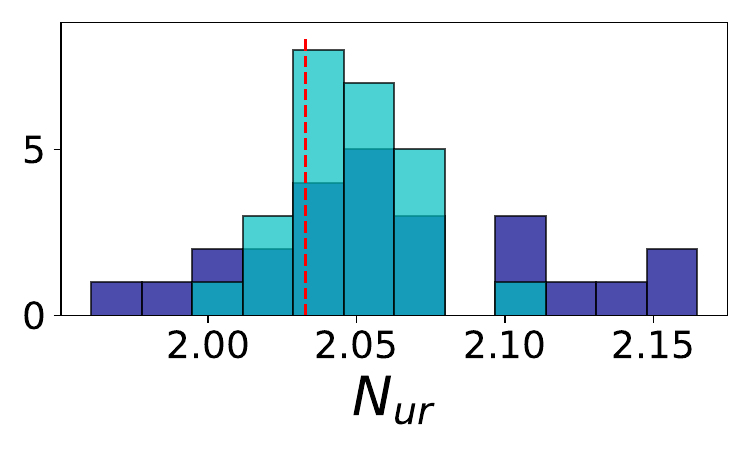}  
  \includegraphics[width=0.195\textwidth]{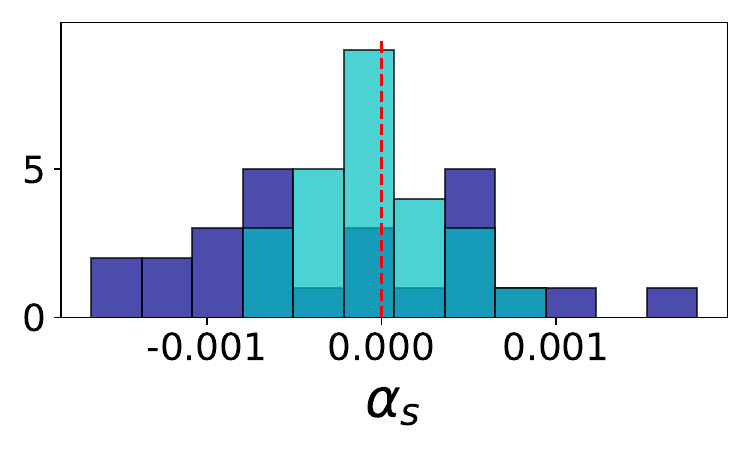}  
  \hfill}
  
  {\hfill
   \includegraphics[width=0.195\textwidth]{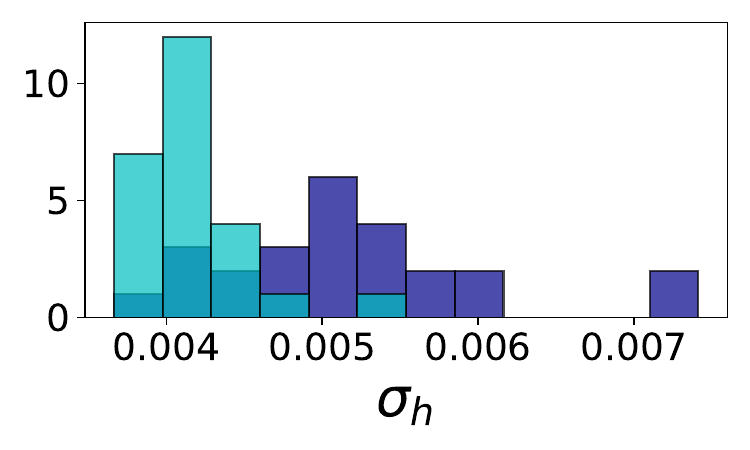}  
  \includegraphics[width=0.195\textwidth]{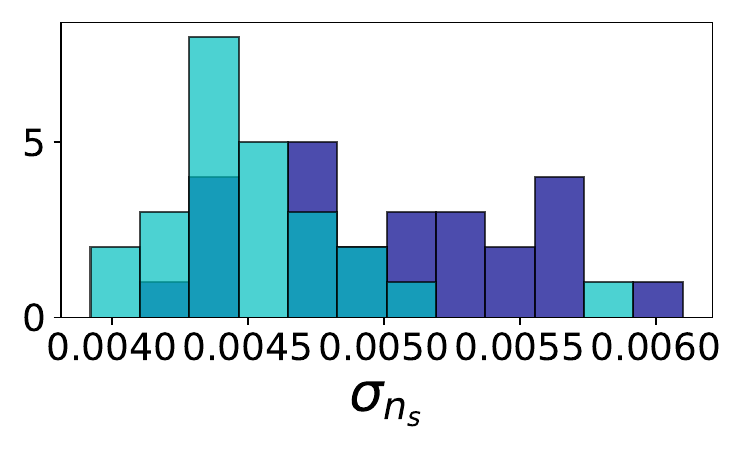}  
  \includegraphics[width=0.195\textwidth]{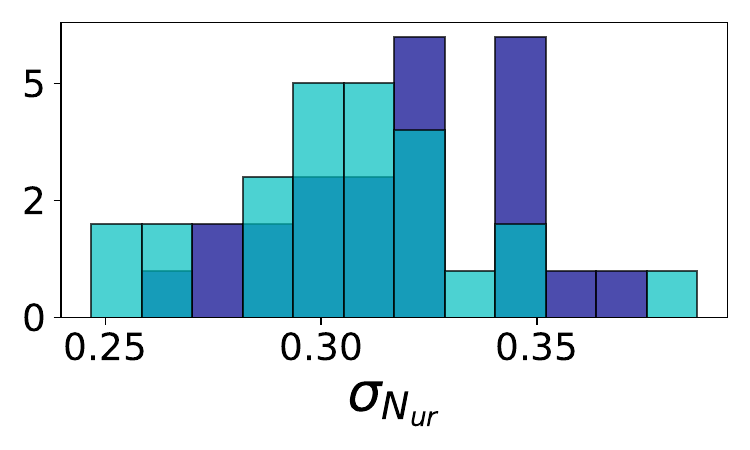}  
  \includegraphics[width=0.195\textwidth]{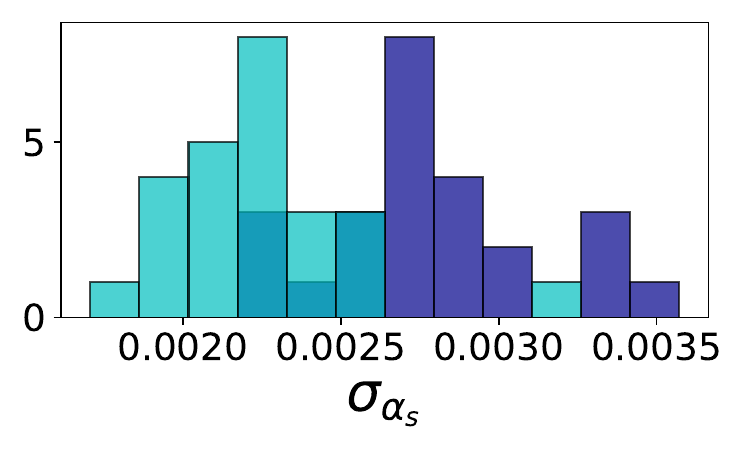}  
  \hfill }
  
\caption{Best fitted cosmological parameters along with the corresponding uncertainties, on the 25 mocks with spherical full sky footprint from $z = 0.15$ to $z = 0.25$, AP distortion, Planck2018 cosmology and the median HOD. Results using the MGP with and without modeling AP distortions are shown in cyan and blue respectively. The true parameters values are pointed by a red dotted line.}
\label{fig:25_ap_cosmo}
\end{figure*}

\begin{figure*}
  \centering
  
  \includegraphics[width=0.195\textwidth]{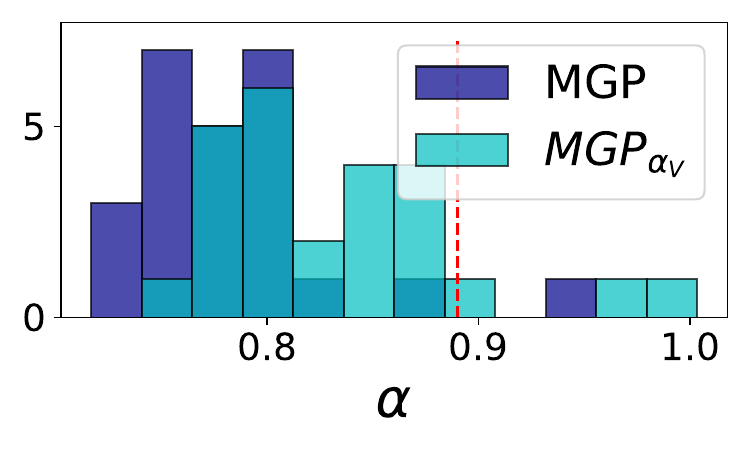}  
  \includegraphics[width=0.195\textwidth]{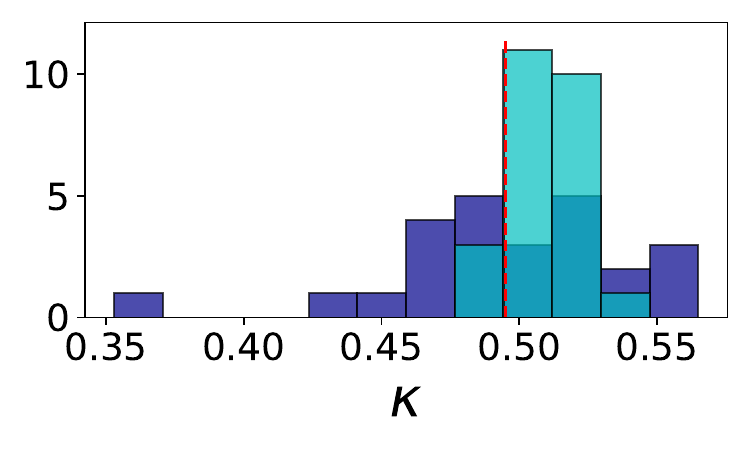}  
  \includegraphics[width=0.195\textwidth]{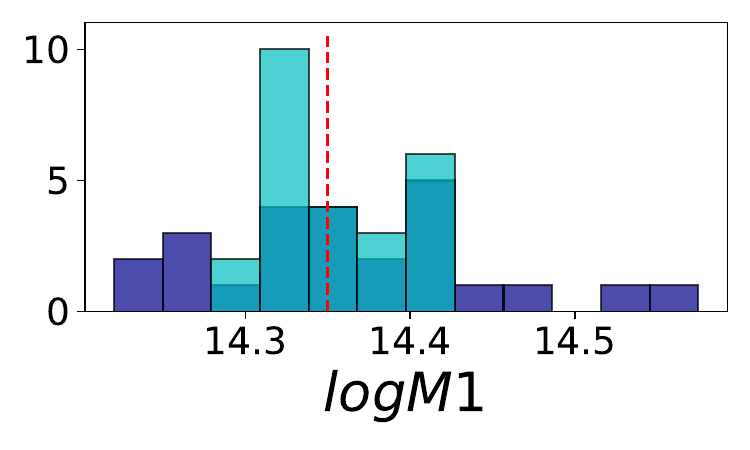} 
    \includegraphics[width=0.195\textwidth]{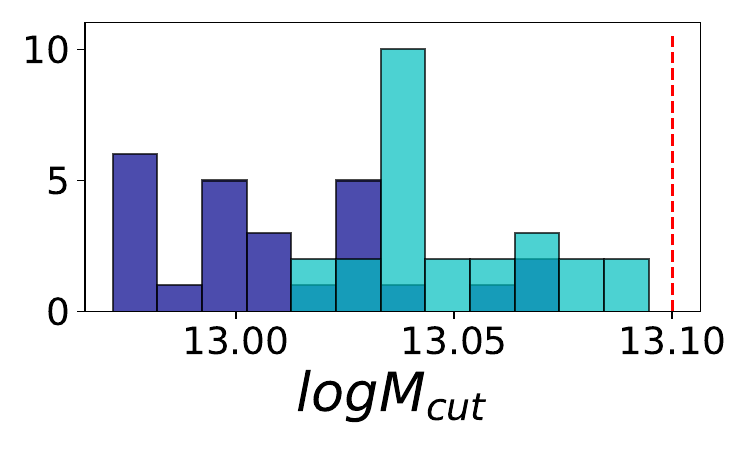}  
  \includegraphics[width=0.195\textwidth]{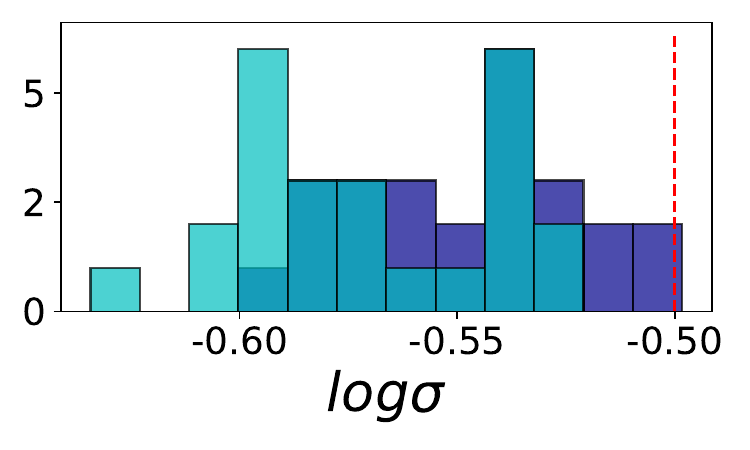} 
  
  \includegraphics[width=0.195\textwidth]{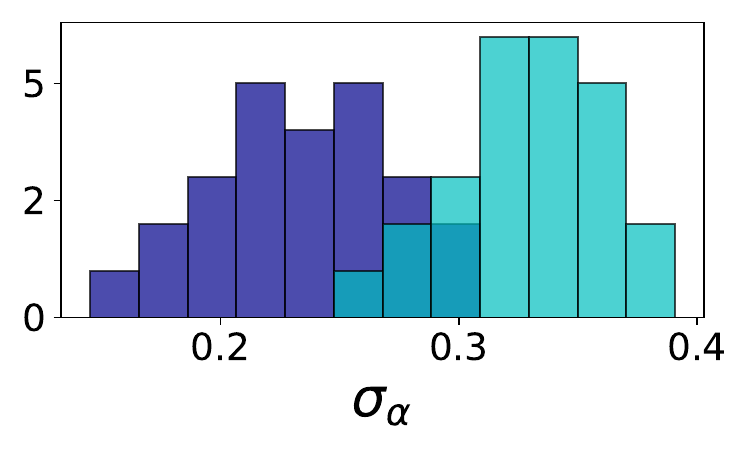}  
  \includegraphics[width=0.195\textwidth]{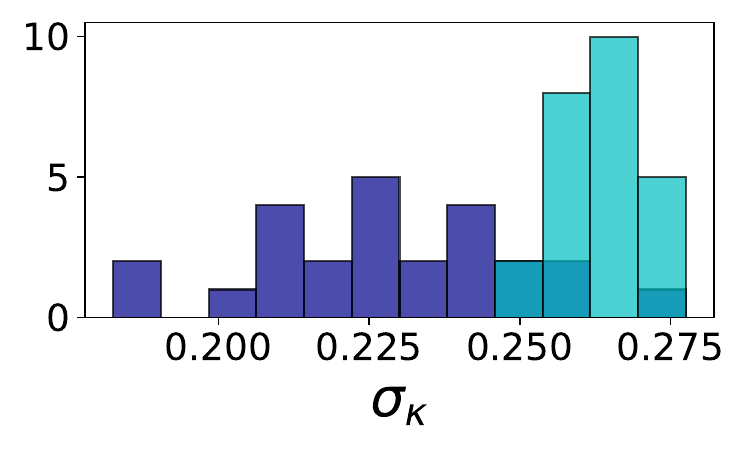}  
  \includegraphics[width=0.195\textwidth]{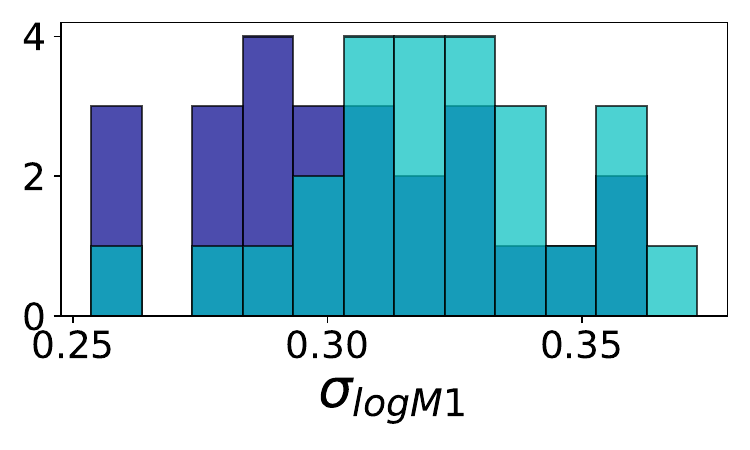} 
    \includegraphics[width=0.195\textwidth]{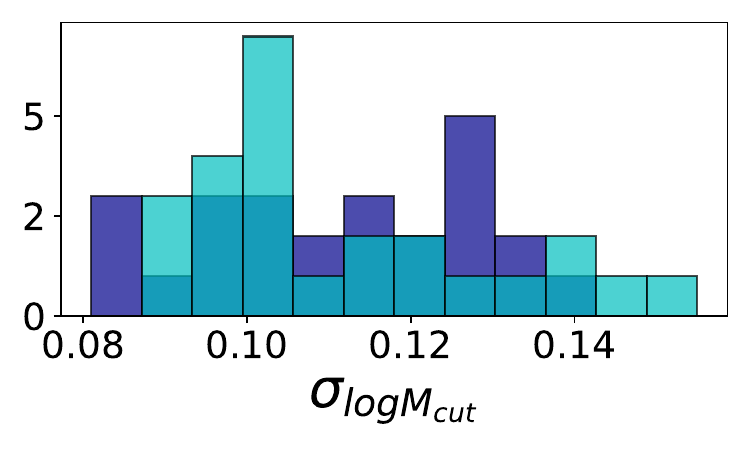}  
  \includegraphics[width=0.195\textwidth]{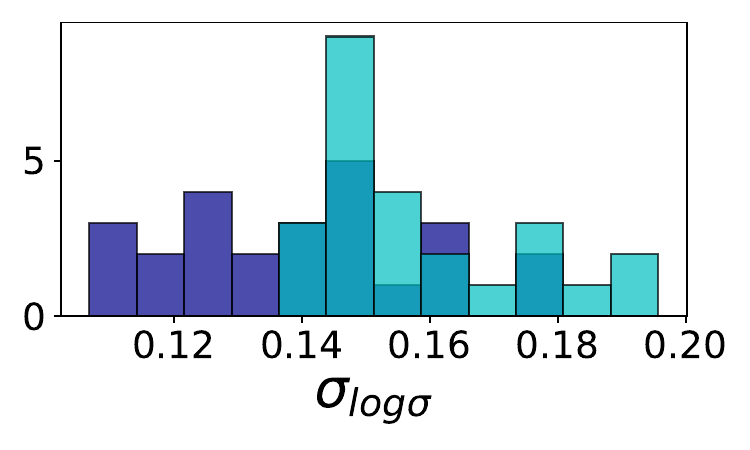}

\caption{Best fitted HOD parameters along with the corresponding uncertainties, on the 25 mocks with spherical full sky footprint from $z = 0.15$ to $z = 0.25$, AP distortion, Planck2018 cosmology and the median HOD. Results using the MGP with and without modeling AP distortions are shown in cyan and blue respectively. The true parameters values are pointed by a red dotted line.}
\label{fig:25_ap_hod}
\end{figure*}

Figure \ref{fig:25_ap_chi2} displays the distribution of reduced chi-squared, best-fit growth rate $f$ and their estimated uncertainties $\sigma_f$.
The best-fit cosmological and HOD parameters with their uncertainties are shown 
in Figures \ref{fig:25_ap_cosmo} and \ref{fig:25_ap_hod} respectively.
Table~\ref{tab:stats_ap} shows summary statistics as in Table~\ref{tab:stats_cubic}, now for the survey-like mocks. 
With a significantly larger cosmic variance, 
the models better fit the data, thus the reduced 
chi-squared distributions are less spread 
and more centered on one compared to the cubic boxes fits. 
The data points are also better described when accounting for AP distortions, the mean of the distributions are $\langle \chi_r^2 \rangle = 0.55 $ for MGP$_{\alpha_V}$ and $\langle \chi_r^2 \rangle =  2.43 $ for MGP for $\mathrm{dof} = (30 - 14) = 16$ degrees of freedom.
In both cases we recover the expected growth rate within one sigma, 
though ignoring the AP effect increases the bias by a factor of ten and gives larger errorbars, with best fitted values for $f$ also more scattered. We also note that compared to the previous cubic box fits, although the cosmic variance of the survey-like mocks is larger, the growth rate is more constrained in both cases.
We see similar results on Figure~\ref{fig:25_ap_cosmo} for the cosmological parameters, when considering AP distortion in the modelling, we reduce the bias and tighter 
the constraints. Except for the dark energy parameters $w_0$ and $w_\textrm{a}$, where although MGP$_{\alpha_V}$ gives tighter constraints, the bias are larger than when using MGP. However, as those parameters are weakly constrained, the difference in the growth rate measurement accuracy between the two models mainly comes from the $\omega_\textrm{m}$ and $h$ parameters.
For the HOD parameters showcased in  figure \ref{fig:25_ap_hod}, we have in both cases unbiased results for every parameters. However as HOD parameters are mostly constraints by small scales where the AP effect is less important, the  MGP$_{\alpha_V}$ model does not bring any substantial improvement, and even gives in average looser constraints then the MGP without AP modelling.

\section{Conclusions}

In this work we reviewed the standard methodology for training an emulator model for galaxy clustering at non linear scales from a suite of dark matter simulations.
We introduced a new Gaussian Process model, allowing to efficiently extend the input parameter space dimension as well 
as account for correlated noise. 


As a proof of concept for our model, we emulated the redshift space 2PCF of galaxies over the input space $X = X_{\Omega} \otimes X_\textrm{HOD} \otimes X_\textrm{s}$ , allowing an additional interpolation over a continuous range of scales, and  making use of the well known correlations between them, both in the signal and the noise.
We implemented our multi-scale Gaussian Process model (MGP) as well as the standard approach which consists in constructing independent Gaussian Processes (IGP) for each separation. 
 We used as a train set 88 cosmologies, 600 HOD and 30 separation scales, all measured in the \textsc{AbacusSummit} suite of n-body simulations.
The input parameter space of the resulting models is composed of 9 cosmological parameters, 5 HOD parameters, and the set of
separations of the 2PCF.

After validating the predictive accuracy of our trained model 
on a test set with 6 cosmologies and 20 HOD, 
we ran inferences using MCMC to check the robustness and
constraining power of our models. 
In a high-precision setting, where we used the average 2PCF of 
25 realisations, ignoring the predicted uncertainties of our GP models
can lead to biased results, specially with the IGP model. 
We also demonstrated that both models are able to recover the expected HOD parameters. Marginalising over such parameters,
we are able to obtain unbiased cosmological constraints. 
Our MGP model returns more accurate and precise constraints on the cosmological parameters than the IGP. 

Moreover, we showed that Alcock-Paczynski distortions can become important compared to the cosmic variance at intermediate scales for the redshifts we are testing. 
Neglecting this effect in an emulator model as it has be 
done in previous work contributes to loosen the constraints 
and increase the bias on the growth rate measurement.

The parameter space of such emulator models could also be
extended to higher dimensions to be able to interpolate 
and marginalize over systematic quantities, extended HOD models,
selection effects, different redshift bins and more. 
Some drawbacks for a too high dimensions training set yet
remain. First, building the training set is expensive. Second, although a Kronecker GP model with a very large number of training points is fast to train, the predictions of the model covariance can be slow .
It is also important to remind that for galaxies at lower redshift, building a model would require more realistic simulations reproducing redshift evolution (light cone), to account for wild angle effects. Applying such model to real measurements would also need beforehand to be validated on realistic mocks using different galaxy-halo connection.

Our MGP model could be used in future works to emulate 
a larger set of galaxy clustering observables and their 
cross correlations such as: galaxy and peculiar velocity auto and cross correlations, galaxy clustering and weak lensing, etc.  Accurate models of these correlations are essential in the era of high precision measurements from surveys such as DESI or Euclid.


\begin{acknowledgements}
      We would like to thank Zhongxu Zhai, Will Percival, Corentin Ravoux for useful discussions.
      The project leading to this publication has received funding from Excellence Initiative of Aix-Marseille University - A*MIDEX, a French ``Investissements d'Avenir'' program (AMX-20-CE-02 - DARKUNI).
\end{acknowledgements}

\appendix

\section{AP effect on IGP fits}
\label{sec:annex:ap}

\begin{figure}
  \centering
 \includegraphics[width=0.8\columnwidth]{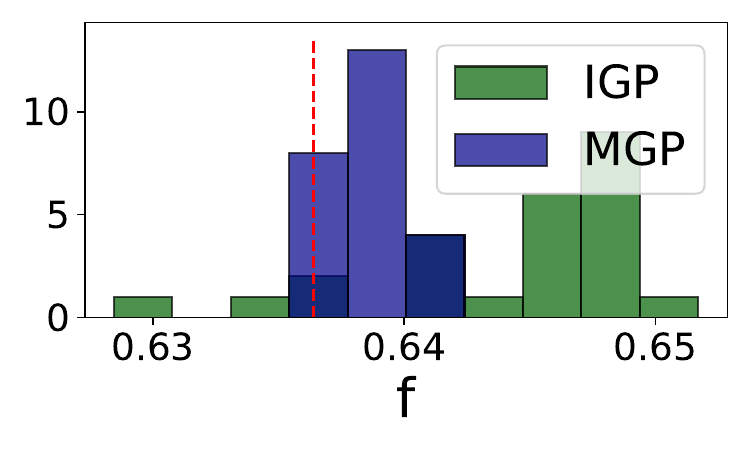}  
  \includegraphics[width=0.8\columnwidth]{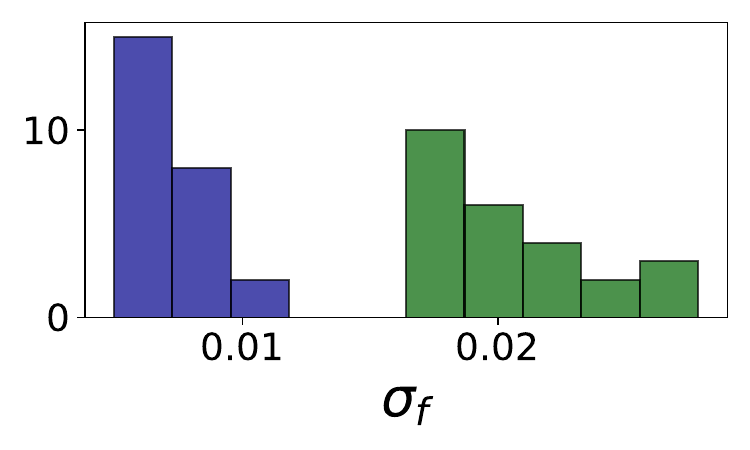}  
  \includegraphics[width=0.8\columnwidth]{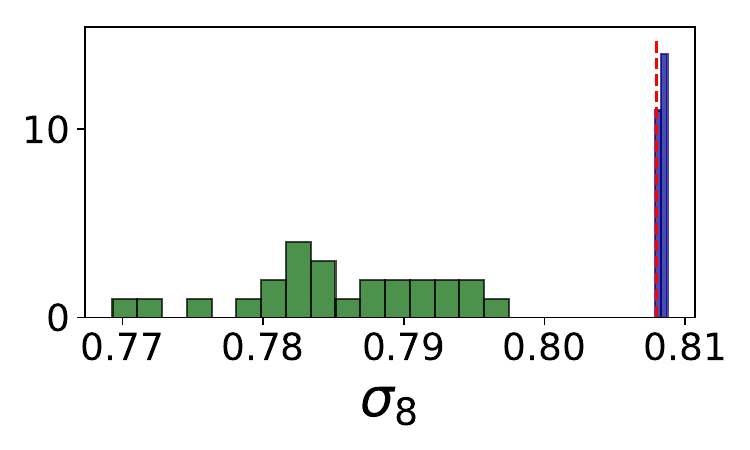}
  \includegraphics[width=0.8\columnwidth]{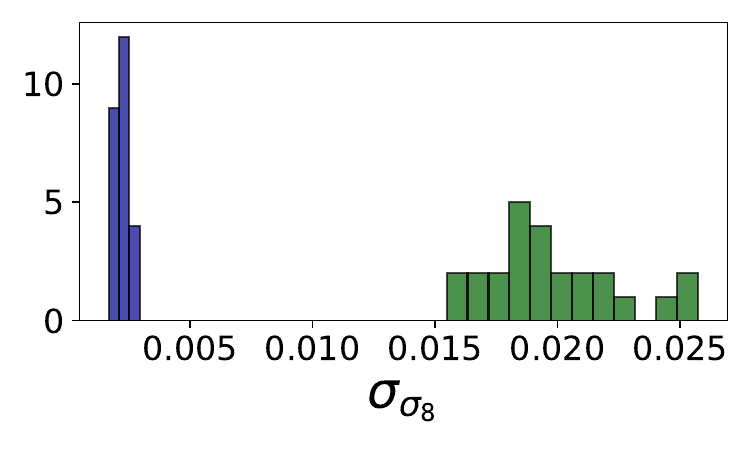}

\caption{Best fitted growth rate and corresponding uncertainties, resulting from the separate fits of the 25 mocks with spherical full sky footprint from $z = 0.15$ to $z = 0.25$, AP distortion, Planck2018 cosmology and the median HOD. Results using the MGP and IGP models are shown in blue and green respectively. The true parameters values are pointed by a red dotted line.}
\label{fig:annex_hist}
\end{figure}

We saw in section \ref{sec:reco:AP} that ignoring the AP distortions can lead to a loss of accuracy and precision, specially for the two cosmological parameters of interest $f$ and $\sigma_8$. Here we compare the impact of ignoring the AP distortions using the MGP and IGP models.
Note that as the IGP is not able to interpolate over scales, the AP effect is usually ignored in standard analysis \citep{yuan_stringent_2022, zhai_aemulus_2022, kwan_galaxy_2023}. While the AP effect could also be modelled using the IGP, it would require to perform an extra approximated interpolation between the trained scales, both for model and uncertainty estimate. 
We fit separately the 25 realisations of the AP distorted clustering, using the both models without modelling the AP distortions. In both cases we rescale the sample covariance matrix with the corresponding spherical volume. 
The the best fitted values for $f$ and $\sigma_8$ along with their estimated uncertainties 
are described in figure \ref{fig:25_ap_chi2}. The best-fit cosmological and HOD parameters with their uncertainties are shown in figure \ref{fig:annex_hist}.
Table~\ref{tab:annex_stat} shows summary statistics for all cosmological and HOD parameters.
We see that when ignoring the AP distortions, although it is not the case for HOD parameters (the very small scales remains unaffected by AP effect),  the MGP model performs significantly better than the IGP for every cosmological parameters. Specially the IGP model gives constraints on $\sigma_8$ biased by more than 1$\sigma$.

\begin{table}
    \centering
    \caption{Summary statistics for the fits of the 25 mocks with spherical full sky footprint from $z = 0.15$ to $z = 0.25$, Planck2018 cosmology and the median HOD. $\langle \Delta_\textrm{p} \rangle$ is the bias, the difference between the true and best fitted value for any parameter $p$. $\langle \sigma_\textrm{p}\rangle$ is the average over the 25 realisations of the (symmetrised) one sigma confidence level.  $\textrm{std}\left( \textrm{p} \right )$ is the standard deviation of the estimated parameter $p$ over the 25 results.}
    {
    \begin{tabular}{l|ccc|ccc}
    \hline
    \hline
      $(10^2 \times )$ & \multicolumn{3}{c|}{IGP}  & \multicolumn{3}{c}{MGP} \\
     $p$ &
    $\langle \Delta_\textrm{p} \rangle$ & 
    $\langle \sigma_\textrm{p}\rangle$ & 
    $\textrm{std}\left( \textrm{p} \right )$ & 
    $\langle \Delta_\textrm{p} \rangle$ & 
    $\langle \sigma_\textrm{p}\rangle$ & 
    $\textrm{std}\left( \textrm{p} \right )$ \\

    \hline
    \hline
$f$ & -0.79 & 2.05 & 0.52& -0.20 & 0.75 & 0.15  \\ 
\hline
\hline
  $\omega_\textrm{m}$  & -0.18 & 0.74 & 0.19 & -0.04 & 0.13 & 0.03\\ 
  $\omega_\textrm{b}$ & -0.01 & 0.11 & 0.02 & -0.00 & 0.05 & 0.01 \\ 
  $\sigma_8 $ & 2.25 & 1.97 & 0.70 & -0.04 & 0.23 & 0.02 \\ 
  $w_0$ & -2.03 & 14.14 & 3.17& -0.03 & 3.27 & 0.41  \\ 
  $w_\textrm{a}$ & 3.25 & 37.54 & 8.42 & 0.34 & 12.68 & 1.67 \\ 
  $h $  & 0.71 & 3.04 & 0.76& 0.15 & 0.52 & 0.09 \\ 
  $n_\textrm{s}$ & -0.01 & 3.2 & 0.72& 0.01 & 0.50 & 0.08  \\ 
  $N_\textrm{ur}$  & -2.46 & 57.32 & 13.49& -2.94 & 31.75 & 5.01  \\ 
  $\alpha_\textrm{s}$ & 0.40 & 1.84 & 0.86& 0.01 & 0.28 & 0.04  \\

    \hline
    \hline

  $\alpha$ & 6.02 & 15.82 & 5.15& 10.54 & 23.24 & 4.75  \\ 
  $\kappa$ & -7.31 & 29.72 & 4.8 & 0.04 & 22.84 & 4.53 \\ 
  $\textrm{log}M_1$  & 0.66 & 12.0 & 4.63& -1.27 & 30.42 & 8.47 \\ 
  $\textrm{log}M_\textrm{cut}$& 4.99 & 6.89 & 2.43 & 8.95 & 11.12 & 2.87  \\ 
  $\textrm{log} \sigma$ & 2.33 & 9.27 & 2.96& 4.74 & 13.97 & 2.56  \\ 
 
    \hline
    \hline
    
    \end{tabular}
    }
    \label{tab:annex_stat}
\end{table}

%
%

\bibliographystyle{aa}
\bibliography{MyLibrary,Biblio}

\end{document}